\documentclass[preprint,12pt,3p,authoryear]{elsarticle}

\usepackage[title]{appendix}

\usepackage{natbib}

\usepackage{amssymb}
\usepackage{subfig}
\usepackage{color}
\usepackage{hyperref}
\usepackage{bm}
\usepackage{amsmath}
\usepackage{multirow}
\usepackage{setspace}
\usepackage{subfig}%[font=footnotesize]
\usepackage{graphicx}
\usepackage{graphics}
\usepackage{array}

\usepackage{titlesec}
\titleformat{\section}[block]{\large\bfseries}{Sec.\arabic{section}}{1em}{}[]
\titleformat{\subsection}[block]{\normalsize\itshape\bfseries}{\arabic{section}.\arabic{subsection}}{1em}{}[]

\begin{document}

\begin{frontmatter}

\title{\large{The emergence of critical stocks in market crash}}

\author[label0]{Shan Lu}
\author[label1,label2]{Jichang Zhao\corref{cor1}}
\author[label1,label2]{Huiwen Wang}

\address[label0]{School of Statistics and Mathematics, Central University of Finance and Economics}
\address[label1]{School of Economics and Management, Beihang University}
\address[label2]{Beijing Advanced Innovation Center for Big Data and Brain Computing}
\cortext[cor1]{Correspondence and requests for materials should be addressed to J. Z. (email: jichang@buaa.edu.cn). }

\begin{abstract}
In complex systems like financial market, risk tolerance of individuals is crucial for system resilience.The single-security price limit, designed as risk tolerance to protect investors by avoiding sharp price fluctuation, is blamed for feeding market panic in times of crash.The relationship between the critical market confidence which stabilizes the whole system and the price limit is therefore an important aspect of system resilience. Using a simplified dynamic model on networks of investors and stocks, an unexpected linear association between price limit and critical market confidence is theoretically derived and empirically verified in this paper. Our results highlight the importance of relatively `small' but critical stocks that drive the system to collapse by passing the failure from periphery to core. These small stocks, largely originating from homogeneous investment strategies across the market, has unintentionally suppressed system resilience with the exclusive increment of individual risk tolerance. Imposing random investment requirements to mitigate herding behavior can thus improve the market resilience. 
\end{abstract}

\begin{keyword}
network science \sep financial system \sep risk contagion \sep market resilience \sep market crash
\end{keyword}

\end{frontmatter}

\doublespacing

%
%\section{Introduction}
%\vspace{-0.4cm}
%\linenumbers

Financial markets are characterized by complex systems which give rise to emergent phenomena such as bubbles and crashes occasionally~\citep{stavroglou2019hidden}.
The price limit for single-security, which usually regards as part of a broader effort to mitigate extreme risk in stock market, has been widely used in China, the US, Japan and Canada, etc. It forbids traders trading stocks at any price above or below a predefined level for the remainder of the day. In Chinese A-share market, for instance, the absolute return permitted is 10\% for every regular stock. Though the single-security price limit is designed to be equal for all securities, it functions as a stock-specified tolerance, and helps protect investors by avoiding sharp price declining or jumping. % instead of a systematic tolerance
Conversely, price limit critics claim that price limit may be ineffective~\citep{bildik2006are,kim1997price} or even feed panic selling in times of market crash~\citep{shen2015china}.
As the traders are in fear of the potential illiquidity when price limit locks their positions~\citep{bildik2006are}, they will collectively sell stocks to seek for liquidity or reduce risk exposure, which in turn smashes market confidence and leads to further downward depression on a wider range of stock prices.
%The price limit thus triggers panic selling that contributes to the emergence of systemic risk which then evolves into system collapse. 
For instance, in the 2015-2016 stock market crash in China, such global sell-off has spread so widely that more than 1,000 stocks prices declining to the daily price limit has become nearly normal for investors. 
From the complex system perspective, if the market confidence is enough to withstand illiquidity shocks caused by the price limits, a system collapse can be avoided. 
How the critical market confidence that keeps the market away from collapse reacts to price limits therefore determines the market resilience, which usually described as the ability of a system to adjust its activities to retain stable when shocks arrive~\citep{gao2016universal}.
In the context, a clear feature of interest is the presence of single-stock price limits' effects on the critical market confidence and system disruption. 

An important depression contagion channel for price limits and market panic is the overlapping portfolios when investors invest in the same equities. As mentioned above, this might be the case if investors sought to hedge their exposure to potential illiquidity of the security that had reached the price limit by trading other stocks in their portfolios. A similar effect might arise if traders who have incurred mark-to-market losses in stocks of price limits face margin calls or are required to reduce their positions to meet the obligation of leverage ratio~\citep{anderson2015resilience,bardoscia2017pathways,gao2018impact}. Additionally, investors may be unwilling to hold securities in fear of that the price limit is driven by information that will affect the value of a wide range of stocks, e.g. the systemic risk~\citep{brugler2018cross-sectional}.
This kind of `loss of market confidence', in particular, plays a non-negligible role in market crisis~\citep{may2010systemic,arinaminpathy2012size}. 
Thus, it is worth exploring the exact knowledge of the relationship between the designed price limit that cause liquidity shocks and the more general `loss of confidence' that can provoke throughout the system. It is also of great importance to probe how the investment behavior and market network structure link together as well as how they influence the association between price limit and critical market confidence, from both theoretical and practical perspectives.

% In face of the portfolio overlapping, we explore the following research questions in this paper: how might the price limits seek to mitigate the initial impact? More broadly, what is the effect of micro-level investment behavior on the relationship between market confidence (as systematic tolerance) and price limits (as individual-specific tolerance) in the depression contagion procedure? 
We explore these by constructing a bipartite stock network. While previous work heavily depends on numerical simulations of networks that are often assumed to be random~\citep{arinaminpathy2012size,may2010systemic,caccioli2015overlapping,caccioli2014stability}, we use the bipartite network constructed by mutual fund share holding data in the real world. We also introduce a contagion mechanism and its analytical explanation on the network for capturing how the market confidence and price limits may contribute to market collapse in times of crisis.

From the theoretical point of view, our approach recognizes some important differences with other complex systems and inherently challenges existing understandings. While previous models have underlined the importance of `superspreader' in ecosystem stability~\citep{arinaminpathy2012size,haldane2011systemic}, our results adversely demonstrate the importance of relatively `small', totally `nested' stocks that drives the system to collapse. Those small stocks, largely originating from homogeneous risk-minimizing investment strategy across the market, determine how the critical market confidence reacts to the price limit. In practice, these findings inject new insights into the market supervision and suggest authorities watch the critical small stocks instead of fully attracted by some systemically important ones.

\section*{Results}
\subsection*{Model Overview}
\begin{figure}[htbp]
\centering
\includegraphics[width= 0.9\linewidth]{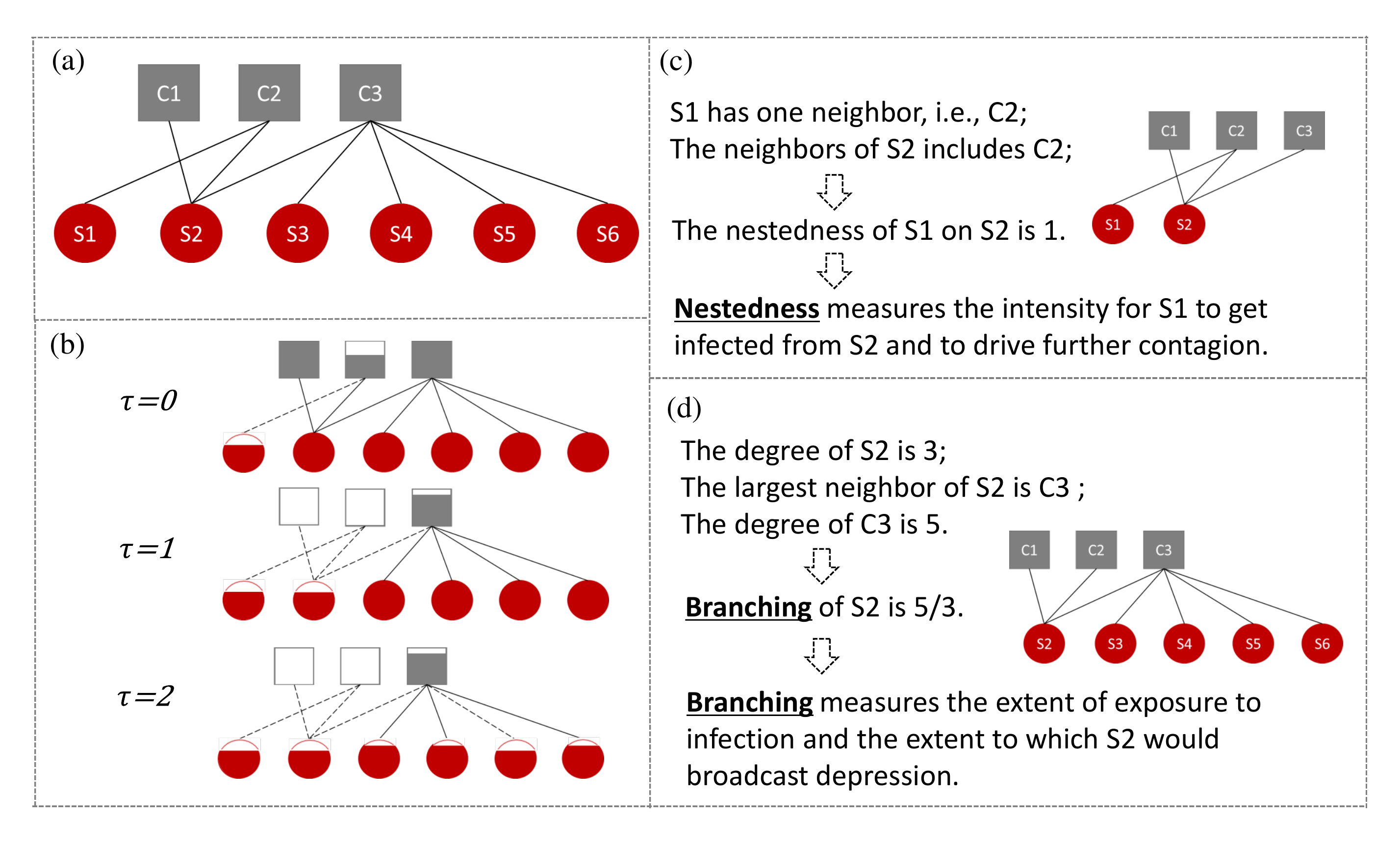}
\caption{{\bf The stock-investor network illustration.} (a) shows the network compositions, where the grey squares are investors and the red circles are stocks. Edges exist only from one kind of nodes to the other and edge weights represent how much the investors invest on the corresponding stocks with respect to market values. It is an undirected weighted network. (b) shows the contagion procedure. The initial shock is stock S1 reaching its price limit and losing $c$ proportions of its market value. The lack of liquidity results from this leads to investor C2 faces a proportion of share holdings being locked. This further makes C2 sell other stocks in hand at $\tau=1$, i.e., S2, which then conveys the depression to C1, and so on and forth. (c) and (d) shows the examples of `nestedness' and `branching' that defined in \textit{Methods}.}\label{fig:example} %section~\ref{sec:analy}
\end{figure}

Here we build a bipartite stock-investor network (see Fig.~\ref{fig:example}). While similar networks have been used to study the financial system stability~\citep{huang2013cascading, delpini2019systemic, poledna2018quantification}, the present paper would mainly focus on the stock market crash and how the network structure determines the association between downward price limit and critical market confidence, which features the market resilience.

Define the absolute value of price limit down as $c$, where $c\in (0,1)$. The market value of stock $i$ at time $\tau$ is $S_{i,\tau}$. For stock $i$, if 
\begin{equation}\label{eq:down_limit}
\frac{S_{i,\tau}-S_{i,\tau=0}}{S_{i,\tau=0}}\leq-c,
\end{equation}
stock $i$ reaches its limit down price, what we called it `failed'. A high absolute value of limit down allows a stock to withstand larger shocks before it is pushed to suffer the lack of liquidity. We study the consequences of shock initially hitting any single stock at $\tau=0$, with the shock taking the form of wiping out a fraction $c$ of its initial values. While $c$ is the limit down threshold, this equals to evoking the stock's failure. An initial failure of a stock that reduces the market values of the investors' liquidation ability will elicit the panic selling on other stocks. 
If the market's demand is less than perfectly elastic, such disposals will result in a short run change in stock price~\citep{cifuentes2005liquidity, coval2007asset}. Subsequently, the externally imposed price limits may dictate additional panic selling which will have a further impact on market prices. See Fig.~\ref{fig:example}(b) for an illustration.

Following the outlined cascading procedure, we integrate the interaction of market confidence into the model as a scaling effect on the liquidity shock. 
Specifically, the $\tau=0$ failure by a single stock results in each of its investors' holding portfolio experiencing a $\tau=1$ shock of magnitude 
\begin{equation}\label{eq:devalue}
\alpha \frac{A_{m,\tau}}{A_{m,\tau-1}},
\end{equation}
where $\alpha$ is the market confidence, $\alpha \in [0,1]$, and $A_{m,\tau}$ is the total stocks' market value held by investor $m$ at time $\tau$. The term $\frac{A_{m,\tau=1}}{A_{m,\tau=0}}$ defines the degree of illiquidity results from the price limits of failed stocks, by which we assume that a depreciating investor may depress the prices of other stocks in holding portfolio according to the relative illiquidity. 
The market confidence $\alpha$ adjusts the illiquidity magnitude by multiplying $\frac{A_{m,\tau=1}}{A_{m,\tau=0}}$ and regulates the market resilience accordingly.
%Its effects on stock prices depression are mild or negligible when $\alpha=1$, but become more severe as $\alpha$ decreases. 
Market illiquidity is linked directly to confidence effects by Eq.(\ref{eq:devalue}). The assumption is that investors who hold the failed stocks could dispose other stocks in their portfolios at lower prices that related to both their liquidity pressure and market confidence. 
This process causes the capital position of other investors holding these same stocks to be eroded. 

We will now have further, $\tau=1$, failures of the stocks connected to the initially infected investors if Eq.(\ref{eq:down_limit}) is realized. This, in turn, may generate $\tau=2$ failures of stocks when these $\tau=1$ failures of stocks convey the price downward pressure to their investors through Eq.(\ref{eq:devalue}). And so on for $\tau=3$ and further. Arguably, the deficiency in the outlined model is that we assume the confidence level remain fixed to be $\alpha$ as the cascade surges through the system. Nevertheless, we believe that it is useful to have a clear understanding of the dynamics of potential system disruption by assuming the confidence level remains universal~\citep{motter2002cascade-based,motter2004cascade}. 
In conclusion, our model captures how the interplay of market confidence and price limits can generate a downward spiral during market crash (see \textit{Methods}). 
More broadly, the critical market confidence that maintains the stability of the system is of primary interest in the presence of risk tolerance of individuals. By unveiling the connection of the two, we obtain the gauging of market resilience in the next section.

\subsection*{Price limits and critical market confidence: the undesirable relationship}

The relevant parameters in the model design are the price limit $c$ and market confidence $\alpha$. The prior interest is the minimum market confidence to guarantee the robustness of the system, or say the critical $\alpha$, given a fixed price limit. Denote it as $\alpha_c$. Here we use dataset of Chinese mutual company's holding positions to establish the bipartite network and use numerical simulations on the real world network to illustrate and clarify the intuition underpinning our model (see \textit{Methods}). 

\begin{figure}[htbp]
	\begin{minipage}{0.5\linewidth}
		\centering
		{\footnotesize (a)}\\
		\includegraphics[width =0.9\linewidth]{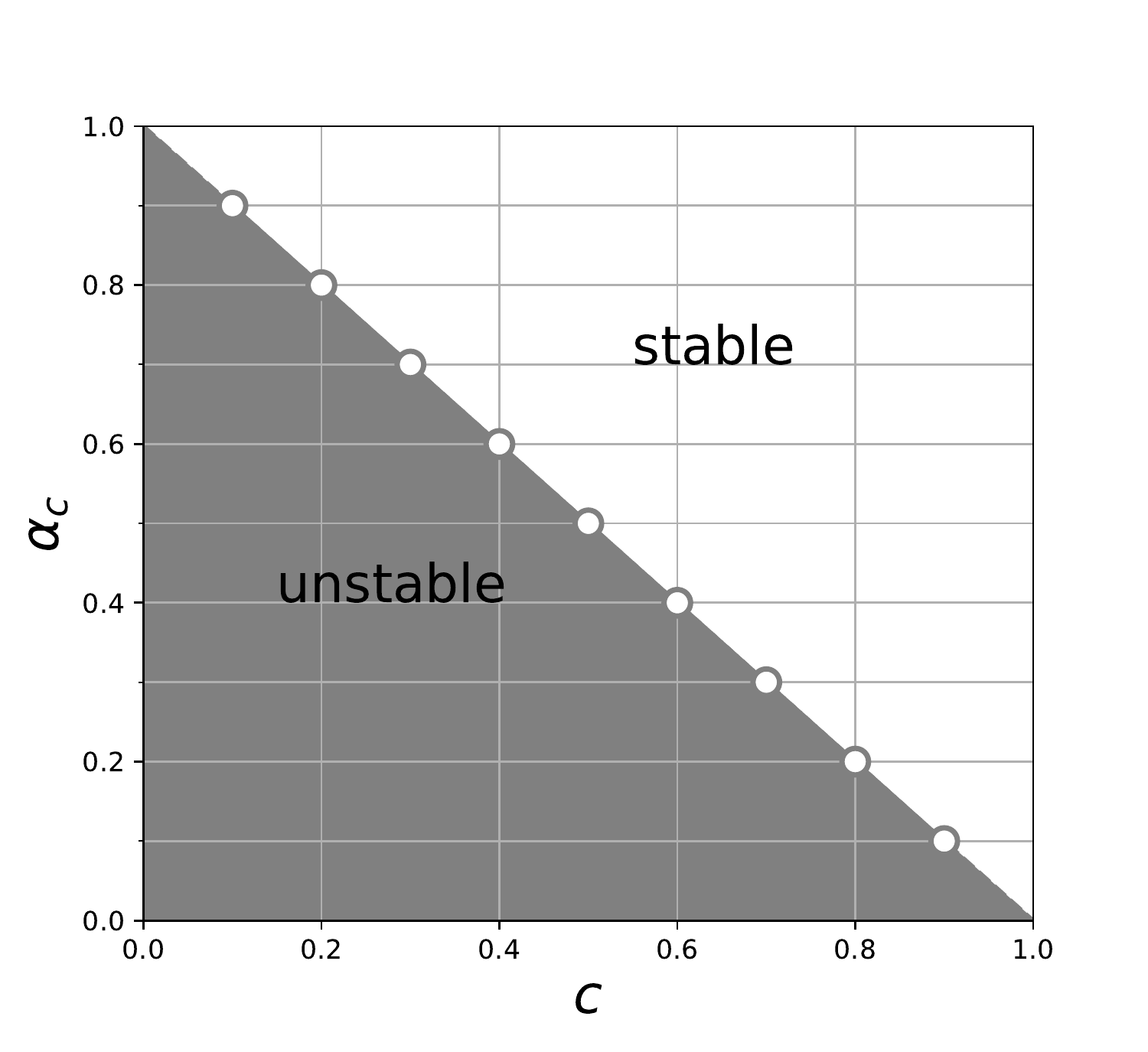}
	\end{minipage}
	\begin{minipage}{0.5\linewidth}
		\centering
		{\footnotesize (b)}\\
		\includegraphics[width =0.9\linewidth]{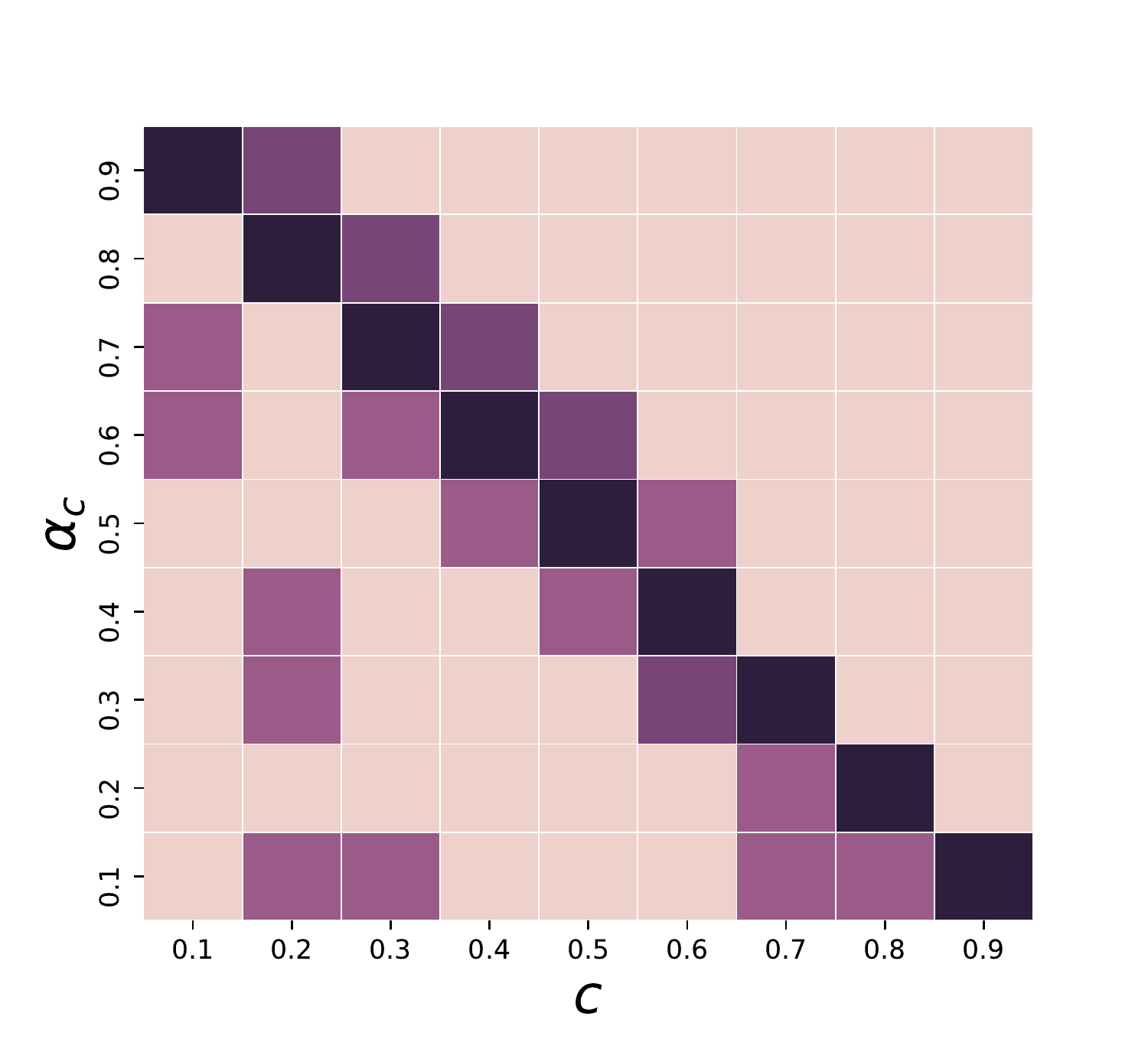}
	\end{minipage}
\caption{{\bf The relationship between critical market confidence and price limit down.} (a) shows the phase transition boundary as $\alpha_c=1-c$, from which the system switches from stable to unstable. (b) shows the ratios of initially shocked stock whose neighbors' maximum $\alpha_{c_i}$ lie in the relative intervals with respect to $c$ ranging from 0.1 to 0.9. The darker the higher of the ratios. The grids on diagnose are the darkest ones, indicating the majority of initial shocked stocks have at least one neighbor with $\alpha_{c_i}=1-c$.} 	
\label{fig:alpha_c}
\end{figure}

To make the result independent on the initial shock, we apply a shock to the stocks one at a time and iterate over the stocks set to obtain the averaged outcomes, see Fig.~\ref{fig:alpha_c}(a). 
%Fig.~\ref{fig:alpha_c}(a) presents the relationship between $\alpha_c$ and $c$. 
The initial shock could, in principle, cause crash of the entire system if $\alpha \leq 1-c$.
The system can switch between stable and unstable, which means that the stock market can either survive and be healthy or completely collapse. More importantly, the phase transition boundary is $\alpha_c=1-c$, indicating that the critical market confidence does decrease with the deepening of down price limit. In other words, the system tends to be more resilient to shocks when the individual risk tolerance is higher. However, as the slope of their relationship is not steep enough as expected, the critical market confidence could not be effectively curtailed with respect to the increase of absolute value of downward price limit. Note that the downward price limit is set as -10\% in the Chinese stock market, where the critical market confidence is still high according to our model. The micro-level market structure that leads to such association has to be further examined, and it is also necessary to investigate the possibility of proper structures in which the critical market confidence can be more efficiently reduced by rising the absolute value of price limits, or the market resilience is to be enhanced.

\subsection*{Driving nodes: the critical ones that prompt system to collapse}
%For an initially shocked stock, we are able to access its stock neighbors' $\alpha_{c}$. %through Eq.(\ref{eq:boundary}). 
As the market confidence $\alpha$ is felt by every investor holding the initially shocked stock and then deliveries the `loss of market confidence' depression to portfolios, it would be stocks with the largest critical market confidence that determined the system-wide critical market confidence, i.e., $\alpha_c$. For every initially shocked stock, denote $\alpha_{c_i}$ as the market confidence needed to avoid its neighboring stock $i$'s failure (neighboring stock $i$ refers to a stock denoted as $i$ that shares at least one common investor with initially shocked stock), whose value at $\tau$ could be accordingly derived (see \textit{Methods}). %by Eq.(\ref{eq:ind_alpha_c}).
Define `driving nodes' as stocks that have common investors with the initially shocked stocks and their $\alpha_{c_i}$ are the largest among others at $\tau=1$ for the initially shocked stocks. One would expect that these driving nodes play the critical roles in regulating system resilience, i.e., the interconnection between the minimum market confidence that needed to keep system stable and the pre-determined price limit.

From the empirical perspective, Fig.~\ref{fig:alpha_c}(b) illustrates the proportions of initial shocks on behalf of their neighboring stocks' maximum $\alpha_{c_i}$ at $\tau=1$. The highest proportions lie on the diagnose of the matrix, demonstrating that it is because the majority of initial shocked stocks connected to at least one of the stocks with $\alpha_{c_i}=1-c$ at $\tau=1$ that drives the system phase transition boundary to be $\alpha_c=1-c$ that shown in Fig.~\ref{fig:alpha_c}(a). 

\begin{figure}[htbp]
	\begin{minipage}{0.5\linewidth}
		\centering 
		{\footnotesize (a)}\\
		\includegraphics[width =0.85\linewidth]{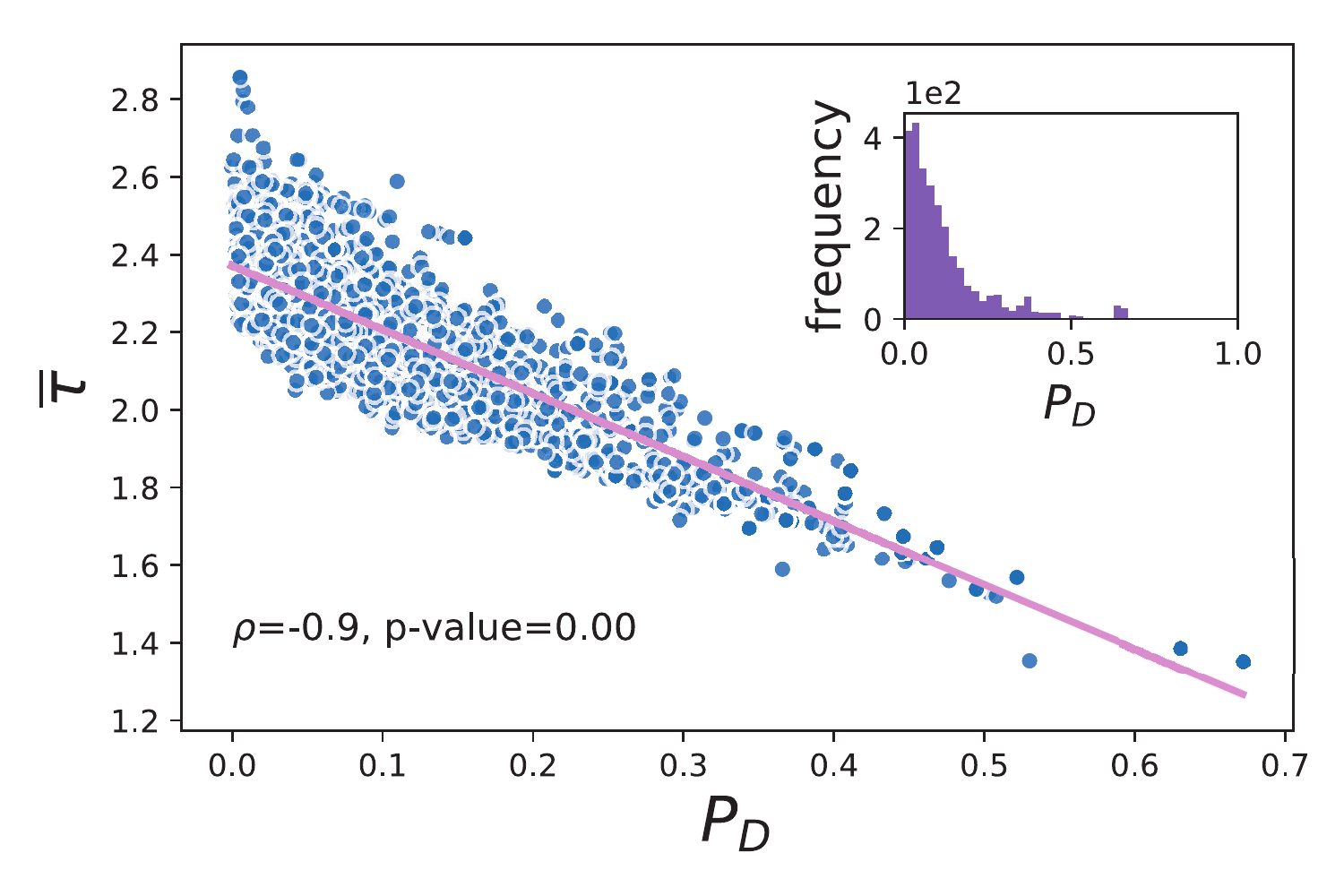}
	\end{minipage}\hfill
	\begin{minipage}{0.5\linewidth}
		\centering 
		{\footnotesize (b)}\\
		\includegraphics[width =0.85\linewidth]{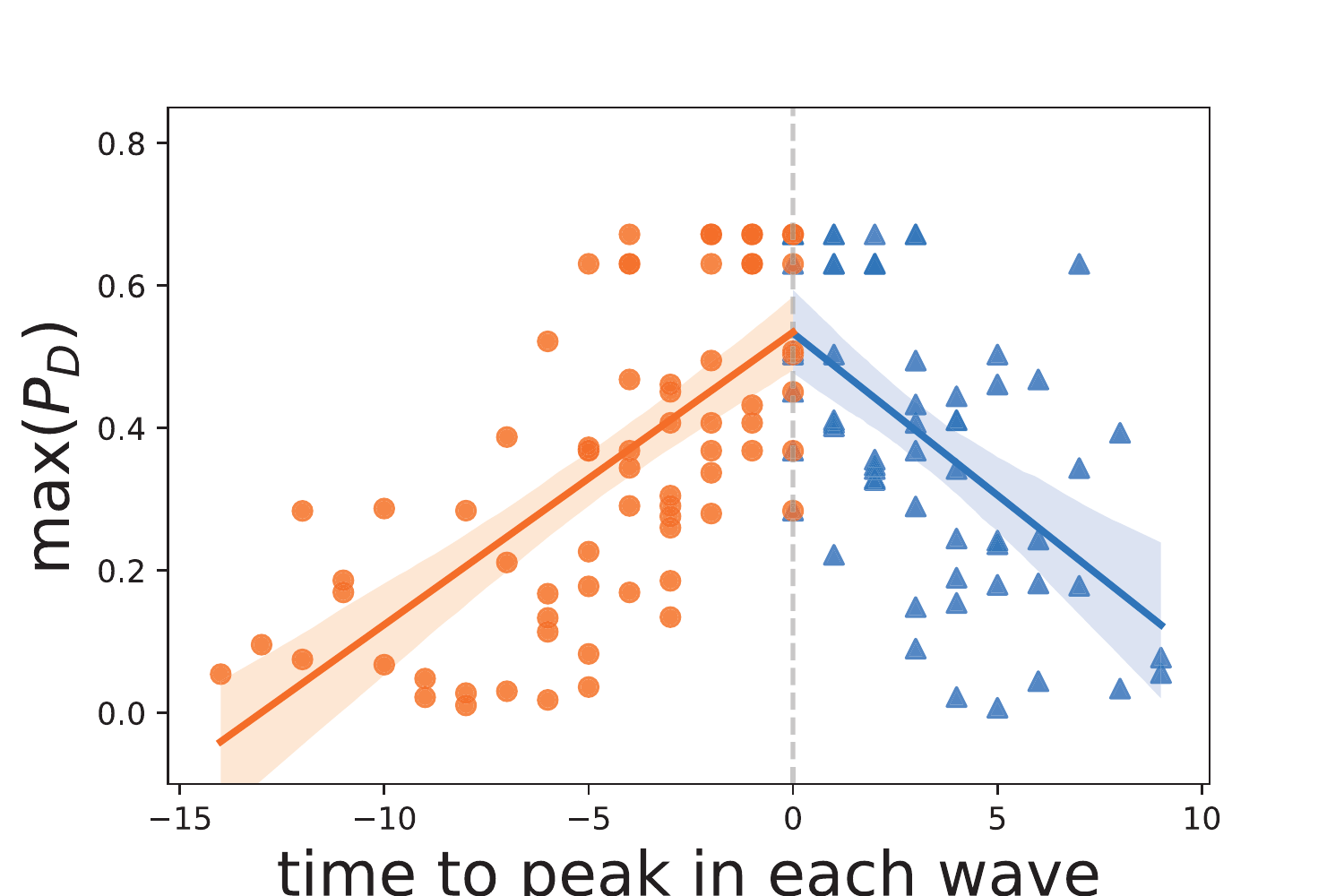}
	\end{minipage}
\caption{{\bf (a) The relationship between $\overline{\tau}$ and $P_D$.} $\overline{\tau}$ is the average cascading steps. $P_D$ is the probability of being driving nodes. The Pearson correlation coefficient and p-value annotated in the plots are for the main axes. The inset in (a) describes the distribution of $P_D$. {\bf (b) The relationship between the time to the peak moments for stocks reaching price limits and the maximum $P_D$ of failed stocks in every minute.} We consider four trading days when market crashes, including June 26, June 29, July 2 and July 3 in 2015, as they considerably speak for the 2015 Chinese market crash and these four days are around the mutual fund ownership data's disclosure date~\citep{lu2018herding}. We first divide each trading day into a couple of non-overlapping time intervals where in each time interval there are stocks reaching down price limits continuously in minute-granularity (see Fig.~S1). These time intervals are called `waves', as they possibly incorporate cascading failure of stocks respectively. We also detect the peak minute(s) in every wave where the number of failed stocks hits the local maximum, indicating a wide range of stocks' failures is happening. The grey dotted line indicates where the peak moments are and the results before or after the peaks are in different colors. 
The Pearson correlation coefficients and p-values annotated are for the points separated by the grey dotted line. }\label{fig:driven_prob_tau_peak}
\end{figure}

Note that the driving nodes are arose from the micro-level network structure, one stock is the driving node of another stock doesn't mean it would be driving node for other stocks. Denote the probability for the stocks to be driving nodes as $P_D$, calculated as the ratio of the stocks' $\alpha_{c_i}$ equal to $1-c$ at $\tau=1$ to all possible initial shocks. Fig.~\ref{fig:driven_prob_tau_peak}(a) exhibits that those of high chances being driving nodes are indeed the ones that fail at the early stage. This coincides with the argument that the success of passing the depression at the start-up phase is the crucial component in cascading failures. Additionally, the inset in Fig.~\ref{fig:driven_prob_tau_peak}(a) indicates that the probabilities of being driving nodes are low for most stocks while a few have high probabilities of being driving nodes. In spite of this, the initial shocks would cascade and cause the stock network collapses provided that there is at least one driving node for any initial attack. Therefore, the nodes with high likelihoods of being driving nodes are critical in determining the system resilience at macroscopic scale. 

On top of the cascading simulation in the bipartite network, the real-cases of market crash also approve the idea that the driving nodes have far-reaching effects on depression contagion. While the contagion model has considered only the case in which there is only one stock failed at the beginning, the initial shock in real market collapse is hard to specify and a more realistic scenario is one in which a network is subjected to simultaneous initial shocks. In fact, when probing the number of newly failed stocks in a minute-granularity manner during market crash, we always find appearance of local peaks (see Fig.~S1). This scenario can be modelled as a sequence of `waves' of newly failed stocks that reaching price limits~\citep{tanizawa2005optimization}. 
And the probabilities of being driving nodes in the bipartite network, i.e., $P_D$, is referred to as the stocks' capability for driving other stocks reaching price limits down in reality. Besides, the contagion model has implied that market collapse would occur in the presence of at least one driving node. Thus we only use the maximum $P_D$ among the failed stocks in each time slot, denoted as ${\rm max}(P_D)$, to qualify the overall driving ability of these failed stocks. 
Interestingly, as shown in Fig.~\ref{fig:driven_prob_tau_peak}(b), the closer to the moment where a wide range of stocks failed, the bigger the ${\rm max}(P_D)$ is. This suggests that failures of stocks with high probabilities of being driving nodes in the established network are indeed capable of leading the market destruction. On the one hand, the results coincide with the simulation results in Fig.~\ref{fig:driven_prob_tau_peak}(a) to certain degree. On the other hand, the results again empahsis the significant influence of these driving nodes to the stability of market, making further examinations of their roles in structure necessary.

%\begin{figure}[htbp!]
%	\centering 
%	\includegraphics[width =0.65\linewidth]{driving_peak}
%\caption{{\bf The relationship between the time to the peak moments for stocks reaching price limits and the maximum $P_D$ of failed stocks in every minute.} $P_D$ is the probability of being driving nodes. Here we consider four trading days when market crashes, including June 26, June 29, July 2 and July 3 in 2015, as they considerably speak for the 2015 Chinese market crash and these four days are around the mutual fund ownership data’s disclosure date~\citep{lu2018herding}. We first divide each trading day into a couple of non-overlapping time intervals where in each time interval there are stocks reaching down price limits continuously in minute-granularity (see Fig.~\ref{fig:four_days_peak}). We call these time intervals `waves', as they possibly incorporate cascading failure of stocks respectively~\citep{tanizawa2005optimization}. We also detect the peak minute(s) in every wave where the number of failed stock hit the local maximum, indicating a wide range of stocks failure is happening. The linear regression fit for both before the peaks and after the peaks are in different colors. } %The definitions of waves and peak could be found in \textit{Materials and Method}.
%\label{fig:driving_peak}
%\end{figure}

\subsection*{Structural roles: small nodes take over and pass on risk}
\begin{figure}[htbp]
		\centering 
		\includegraphics[width =0.6\linewidth]{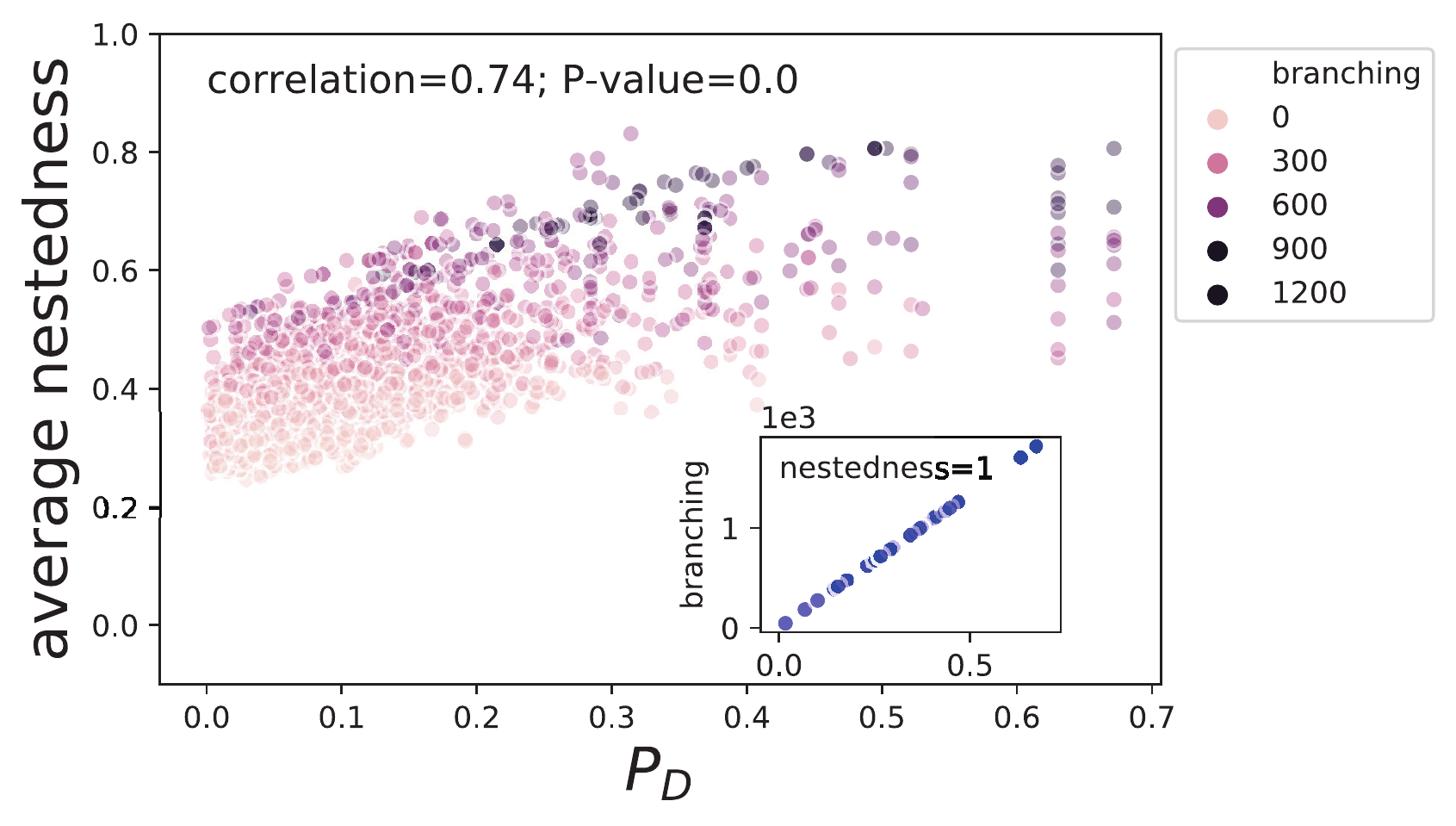}
\caption{{\bf The relationship between average nestedness and $P_D$.} $P_D$ is the probability of being driving nodes. The definitions of nestedness and branching could be found in \textit{Methods}. The correlations annotated in the plots are for the main axes. The inset illustrates the relationship between $P_D$ and branching for stocks with the average nestedness equal to 1. }\label{fig:nestedness_branching}
\end{figure}

In knowledge of the importance of driving nodes, how they emerge from the investing activities is the major concern.
From the theoretical perspective, the driving nodes, or say, stocks with $\alpha_{c_i}=1-c$ in the present real data case, originates from completely sharing neighbors with initially shocked stocks (see Eq.~(\ref{eq:ind_alpha_c}) in \textit{Methods}). %other network may not because ep2 may not equal to 0
And the probability of being driving nodes are further determined by the exposure to connections with initially shocked stocks. 
Thus we define nestedness as the ratio of overlapping investors and branching as the degree of a stock relative to its largest investor's degree (see Fig.~\ref{fig:example} and \textit{Methods}). In general, the nestedness of one stock on a specific initially shocked stock measures how likely it gets infected by the initial shock and reluctantly becomes driving node to pass on depression contagion. The branching of a stock unfolds the balance between its variety in terms of number of neighbors and the diversity of its largest neighbor' investment portfolio. Therefore, branching measures the potential of a stock for being driving node if other stocks among its neighbors' portfolios got shocked.
Matching the driving nodes' probability $P_D$ with nestedness and branching, we find that the odds of being driving nodes are positively correlated with the other two, see Fig.~\ref{fig:nestedness_branching}. On one hand, the stocks which have high nestedness are doomed to have high probability of being driving nodes. On the other hand, the stocks which are of high nestedness tend to possess high level of branching. The $P_D$ of stocks that nestedness equal to 1, in particular, are strictly proportional to their branching. Note that nestedness equal to 1 indicates that all of the stocks' nearest neighboring stocks share completely same neighbor(s) with them but have more neighbors than them. Thus the perfect linearity in the inset of Fig.~\ref{fig:nestedness_branching} reveals that the probability for these stocks to be driving nodes depends on their branching, that is, how their investors diversified portfolios. In particular, those of high probabilities of being driving nodes are connected to investors holding a wide range of equities.
By being so, the highly nested stocks are the most likely to absorb the depression risk from their influential neighbors at the early stage and broadcast shocks to other branches of the system. 

Moreover, we find that those driving nodes are mainly small-cap stocks, with low degrees but high branching and nestedness (see Fig.~S2). The larger-cap stocks, on the contrary, connect to more mutual fund companies, and thus have lower nestedness and branching than the smaller-cap ones. The large gap in average degrees between the large-cap and small-cap stocks implies that while the large-cap stocks are popular among all the mutual fund companies, the small-cap ones could only attract a few mutual fund companies. Additionally, Fig.~S3(a) shows that a large proportion of stocks are of small degrees whereas a few stocks are held by almost everyone of the mutual fund companies. Unlike the stock degree distribution, the mutual fund companies overall have high degrees, in particular a few mutual fund companies hold nearly two thirds of the listed stocks. Therefore, aiming at minimizing the investment risk, investors like the mutual fund companies who hold large numbers of stocks tend to invest on the popular stocks (large-cap) and the unpopular ones (small-cap). The popular ones become severely overlapped but the unpopular ones do not. This leads to the small-cap stocks become the nodes of high nestedness and branching in the network, making them relatively `small' when compared with their neighboring stocks but become the driving nodes that could turn over the system.

Different from previous studies in which the importance of `super-spreader' in network is emphasized~\citep{arinaminpathy2012size,haldane2011systemic}, the present fact that driving nodes could largely be recognized by nestedness and branching prompts the idea of nodes with few neighbors but having one important neighbor would play essential part in our story. In other words, the connections to investors that having exposures across a wider set of stocks make the small stocks possess higher risk on taking-over the depression. Stocks that are nested too much on others should be protected first for the sake of the whole system. And the reason behind this is the homogenous investment strategy that seek for not only wide diversity but also preference on some particular stocks for their safety (see Fig.~S4).

Many different mechanisms have been suggested in the literature to account for such a high degree of similarity across portfolios including connections between mutual fund managers and corporate board members, herding behavior and imitation of successful diversification strategies~\citep{cohen-cole2014trading, corsi2016when}. Another potential reason behinds the similarity of investing pattern is the investing concentration on a range of stocks with high social trust in Chinese stock market as it is believed that stocks with high social trust have smaller crash risks~\citep{li2017trust}. These prudential investment strategies are designed to enhance the market resilience for shocks. However, they lead to a more densely connected heterogeneous financial market and the emergence of small but critical stocks that take over initial shocks and drive further depression, thus undermine system resilience.

\subsection*{Risk contagion: cascading patterns due to driving nodes}
\label{sec:real}
\begin{figure}[htbp]
\centering
	\begin{tabular}{@{}m{6cm} m{5cm} m{5cm}@{}}
	\multirow{2}{*}{ \includegraphics[width= 6cm ]{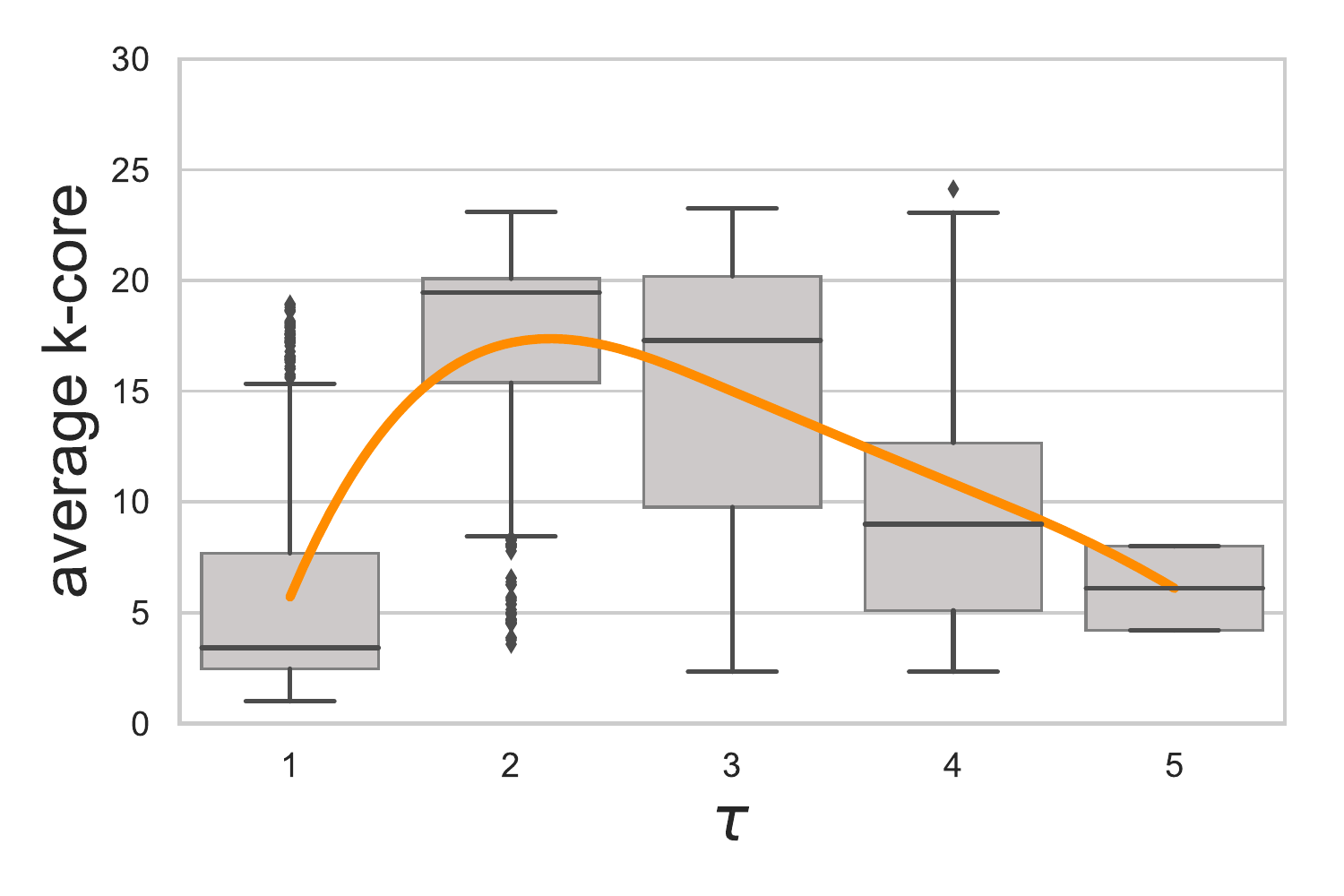} } &
	\includegraphics[width =5cm]{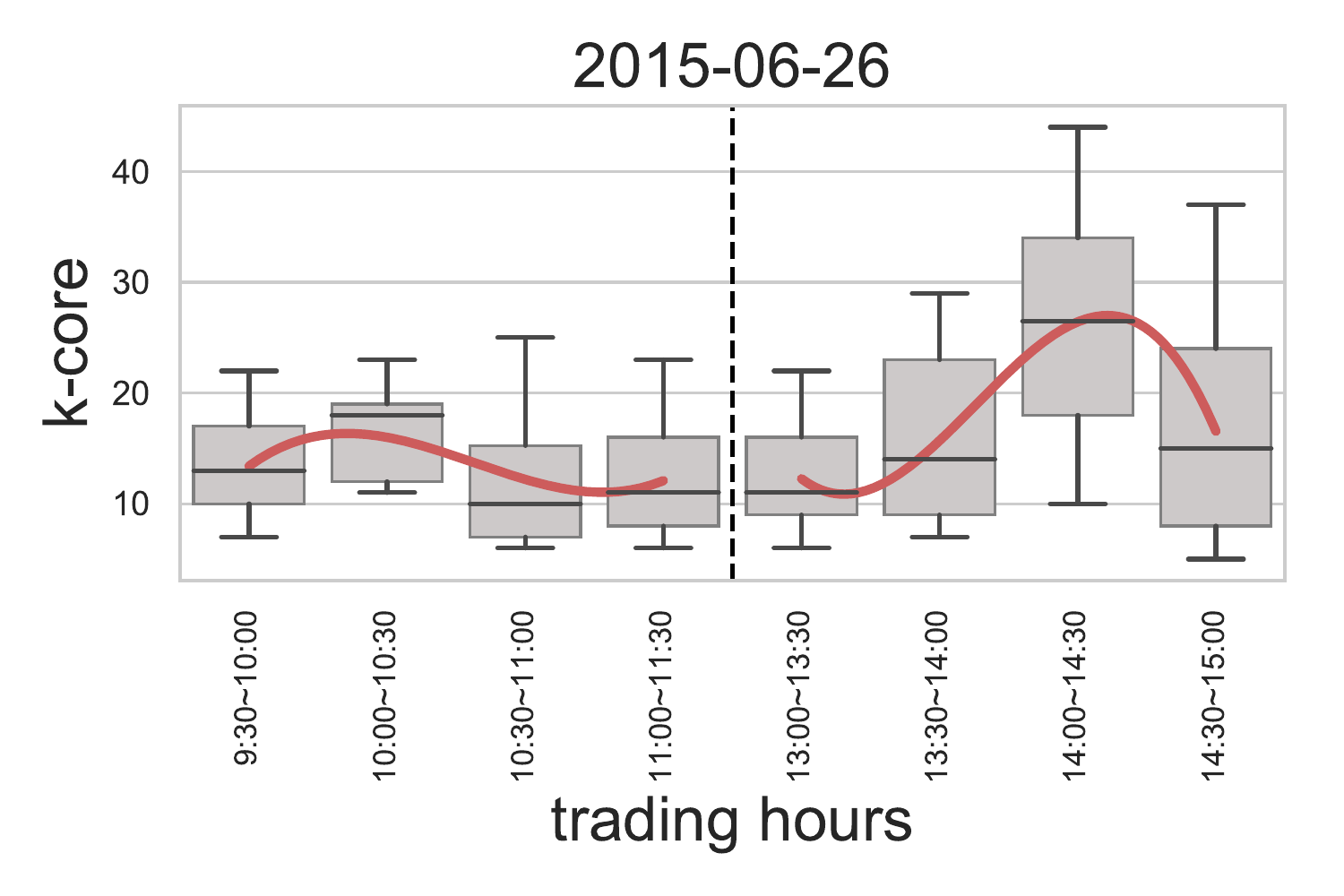} & 
	\includegraphics[width =5cm]{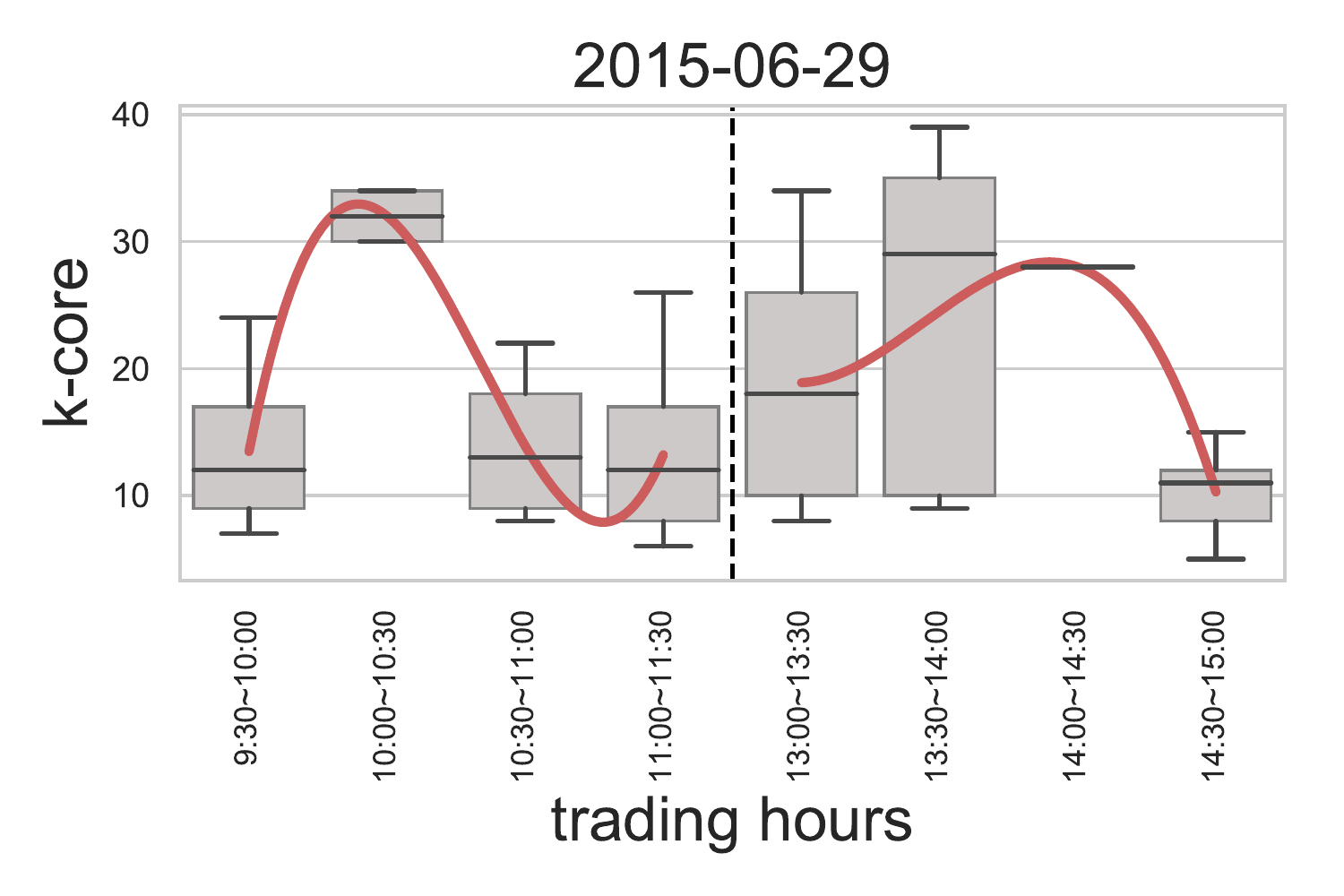}  \\
	& 
	\includegraphics[width = 5cm]{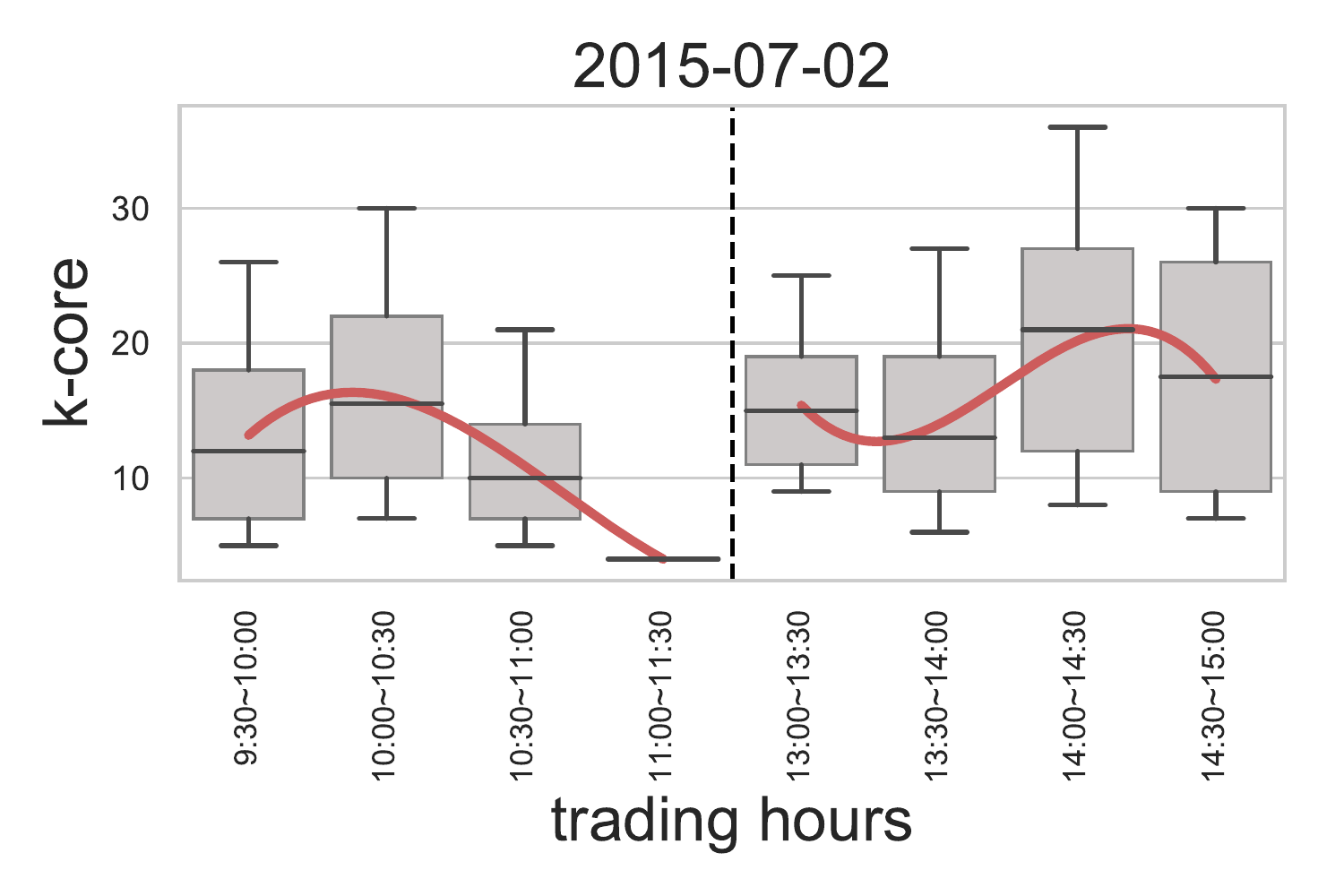} & 
	\includegraphics[width = 5cm]{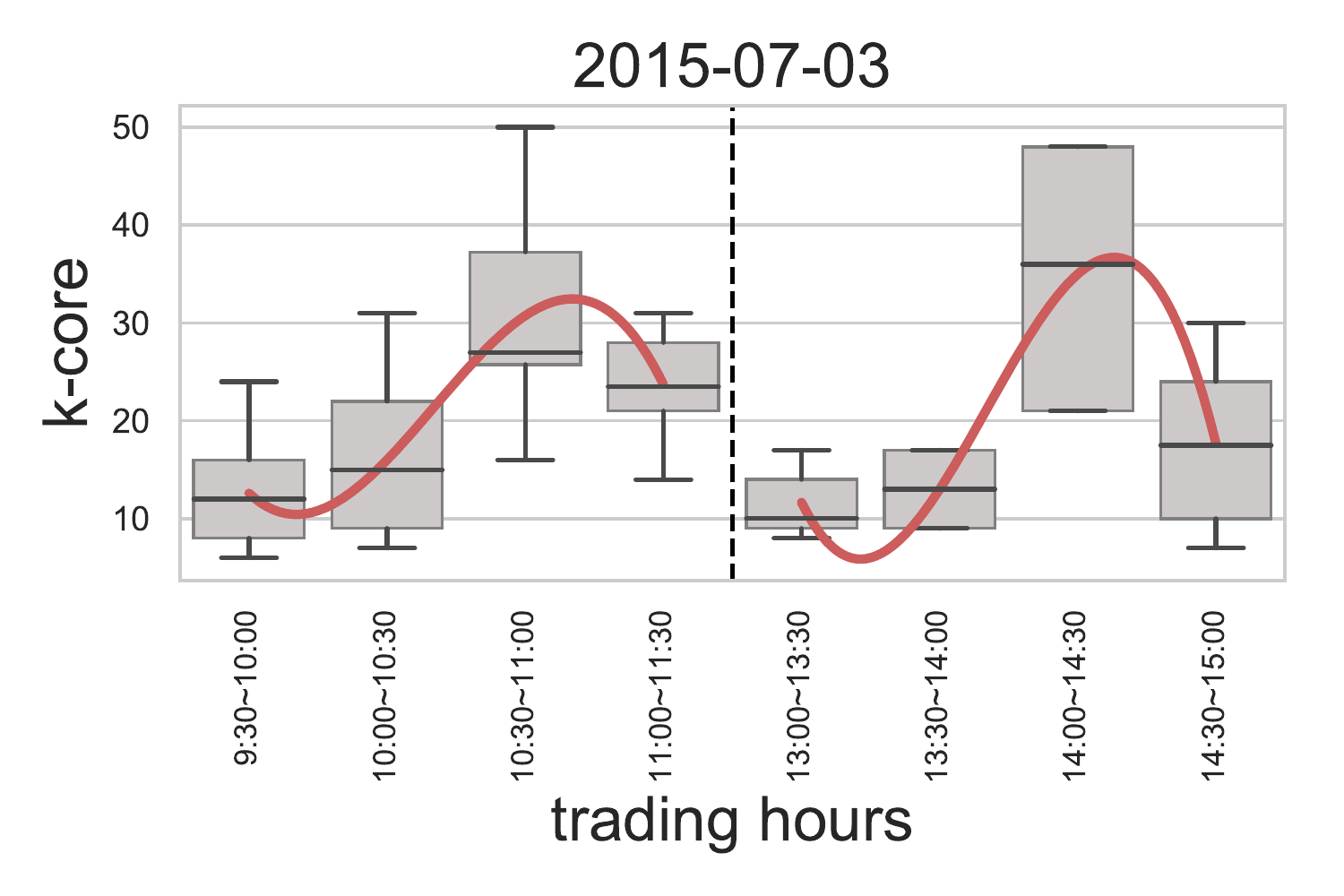} 
	\end{tabular}
\caption{{\bf The $k$-core index of stocks reaching the limit down prices.} The left panel presents the averaged $k$-core index at each $\tau$ in the simulation course from the proposed contagion model. The boxes show the distribution of the averaged $k$-core index at $\tau$. The smooth line (orange) shows the trend of the mean of the averaged $k$-core index at each $\tau$. The right panel which contains four subplots shows the $k$-core indexes of stocks reaching the limit down prices in four market crashing days. The smooth lines (red) depict the trend of the mean of the stocks' $k$-core indexes. The grey dotted lines imply the lunch breaks. For the purpose of clearly visualizing data, outliers have been dropped.} \label{fig:4days_k_core} %note the simulation is averaged over averages and the real cases are just average.
\end{figure}

%Next we explore the extent to which the cascading of the contagion model coincide with real-world cases, as it would suggests, in part, whether our network definition and contagion model is effective. 
Considering the particularity of driving nodes in the microstructure and their critical roles in the acceptance and diffusion of depression, they may lead to a formation of stable macroscopic cascading patterns. 
Understanding such patterns not only helps to recognize the systemic impact of driving nodes, but also offers references for precautionary actions of collapse prevention and even brings about the power of prediction. 
Given the fact that the stocks of high probabilities of being driving nodes are those of high nestedness, a reasonable path for risk contagion would be from the network periphery to the core and then spread to the entire system. Here we use $k$-core index as the description of the nodes' locations in network for its effectiveness in detecting cores and peripheries~\citep{kitsak2010identification}.
The left panel in Fig.~\ref{fig:4days_k_core} clearly demonstrates that the initial attack towards the system first hit the periphery, where the driving nodes locate, and then spreads to the inside. By then, a wide range of failures emerges, propagates to the whole network and results in the system collapse. Note that the contagion dynamics are not long-lived, as the simulation always terminates within a few steps, due to the fact that our network is a rather small one (see Fig.~S5 and Fig.~S6 for how the cascading proceeds). 
%However, we emphasize that time does not appear to be an explicit variable in the sequence of cascading. 

We also probe the $k$-core index distribution using the real-world stocks' failing procedure and find great similarity, see right panel of Fig.~\ref{fig:4days_k_core}. A similar trend of the contagion procedure is found in these four market crashing days, separating by the lunch breaks. To be specific, the order of stocks reaching their limit down prices in the real world is similar: from outside to inside, then from inside to outside (see also for Fig.~S7). This validates that our model approximation gives agreement similar to that seen in real market crash. 

More importantly, one insight in addition to a previous study, that the root case of the system collapse is the extinction of nodes located in the maximum $k$-core of the network~\citep{morone2019k-core}, has emerged. We argue that the driving nodes on the periphery are essential, as they firstly pass on the failures to the core nodes inside the network, that lead the system reach global failure eventually.

Admittedly, the results could only provide slim prediction power of stocks failures due to the fact that the network is built on the mutual funds investors while the individual investors are the majority traders in Chinese stock markets. However, the network is still of valuable representation for the whole market considering that individual investors are easily allured by mutual fund institutions' holding positions and investment trending and that mutual funds occupy a large fraction of overall trading value in mature markets like US, Japan and Hongkong~\citep{lu2018herding}.
More importantly, the success on approximating the real-world stocks' failing process at the macro-level endows the contagion model with an early warning capability. And the results also suggest that the small stocks, which play the roles of driving nodes on the network periphery, should be protected or isolated for precautionary purposes.
Overall, the contagion model would be good a resemblance to the actual market crash. 

\section*{Discussion}
Though the single-security price limit is widely used in stock markets of different counties, the exact knowledge of its influence on market resilience is still unknown. 
In the proposed bipartite network based on common asset exposure in stock market, we specifically study the relationship between the critical market confidence $\alpha_c$ that could maintain system stability and single-security limit down $c$ in a risk contagion model design. The linear relationship between the two, which signals the verge of the financial system becoming unstable, implies that $\alpha_c$ cannot be reduced significantly when rising the absolute value of $c$. 
The fine-tuned price limit, in essence, cannot drastically alter the critical market confidence as expected. From this perspective, the slope of $\alpha_c$ and $c$ would be a new indicator to reflect the system resilience. 
The results are similar if we use mutual funds as the investors in the network instead of mutual fund companies with respect to network structure and the embedded relationship between critical parameters (see Fig.~S8 and Fig.~S9). This sheds lights on the counterproductive of the accordant single price limit setting in the circumstance of investment behavior pattern like China.

Essentially, the verge of the system stability is awarded by the overall similarity among investments. Even though fund managers are professional investors whose diversification strategies cannot be reduced to random selection of assets, several studies have mentioned that the investing behavior among mutual funds is similar~\citep{coval2007asset, lu2018herding,delpini2019systemic}. The herding behavior is more severe in Chinese stock market where we find that similar heterogeneity characterizes the stocks: most stocks are found in the portfolios of a few funds, but some stocks enjoy huge popularity and are held by almost every fund. Stocks with investments from few mutual fund companies are mostly held by those who tend to over-diversify their portfolios. Therefore, they enjoy high nestedness and branching, indicating that they are relatively small in terms of number of investors but have severe portfolio overlapping with other stocks and are able to link the initially shocked stock with other stocks through these investors. On the one hand, they can quickly fail in reaction to the losses of the initially shocked stock's illiquidity even with considerable level of price limit. 
On the other hand, they increase the chances that other stocks will be exposed to investors who experiencing panic selling in the first round of depression contagion, acting like driving nodes for further system collapse. 
That is to say, when the failure of such a stock triggers contagious illiquidity because of price limit, a large number of its investors' linkages also increases the potential for contagion to be extremely widespread. 
In conclusion, the small stocks are vulnerable to the first-round of failure and are critical to the further round price limits implements and illiquidity propagation. 
Additionally, knowing that these small stocks are usually blind spots for regulators, our results highlight their critical roles in determining the system resilience in reaction to individual risk tolerance. And it is the herding behavior on diversification strategies that leads to the dominance of small but critical stocks in the system.

\begin{figure}[htbp]
\centering
\includegraphics[width =0.65\linewidth]{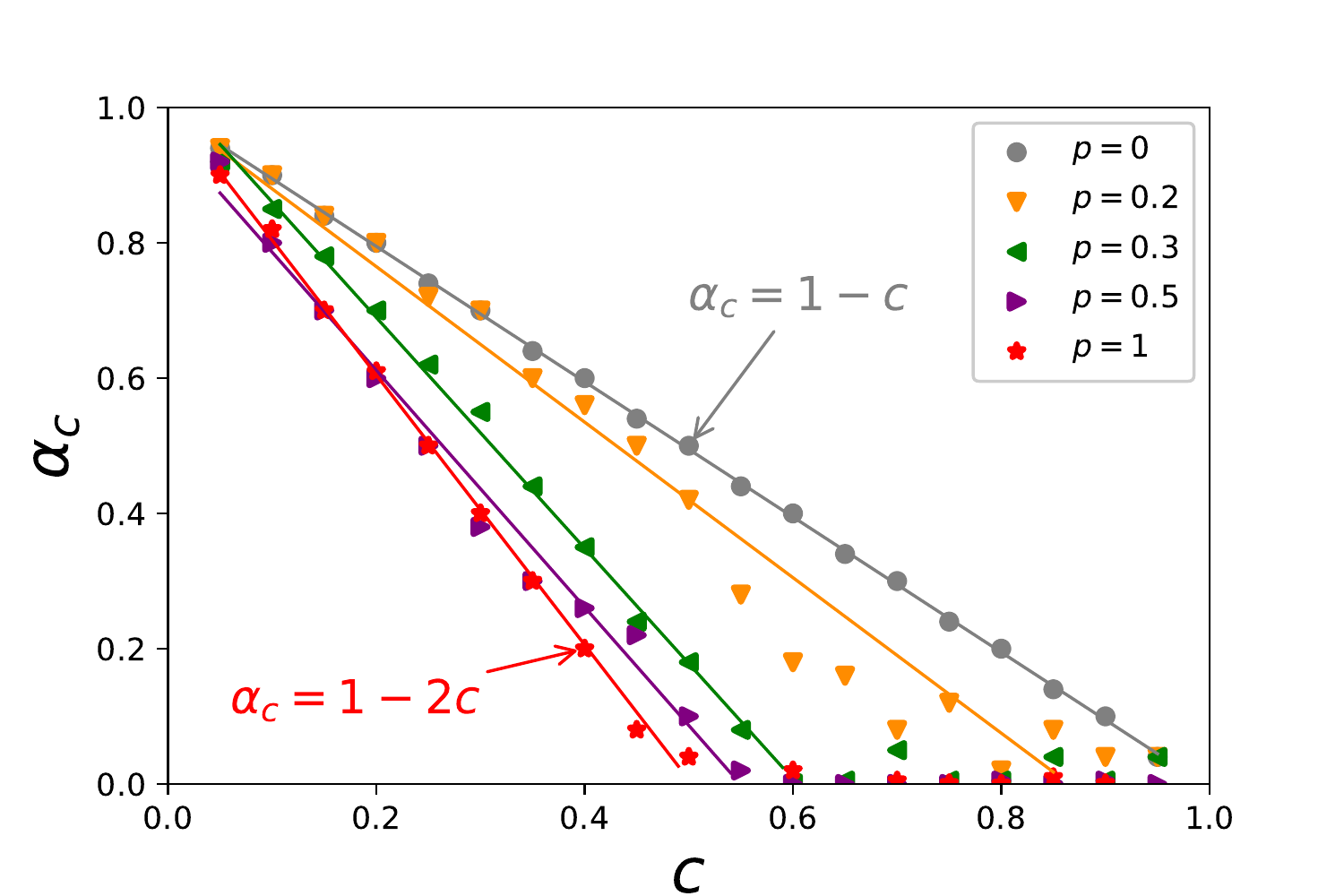}
\caption{{\bf The relationship between $\alpha_c$ and $c$ when randomizing the stock-investor network.} $p$ denotes the proportion of which the edges in the original network are randomized. The results for different $p$ are shown in different colors and marker shapes, along with lines of best fit (results with $\alpha_c \approx 0$ are neglected in the lines of best fit). 
For each $p$, we delete a proportion of $p$ edges in the original network and generate the same amount of edges apart from the $1-p$ edges that have been kept in the network. Thus, $p$ indicates the extent of randomization. When $p=1$, for example, random bipartite networks that possess the same amounts of nodes and edges with the real network are generated. On the contrary, when $p=0$, there is no randomization and the original network is completely kept, which will give us the results in Fig.~\ref{fig:alpha_c}(a).
All the edge weights are set to be equal in the randomization experiments for simplicity. The initial attacked stocks are also randomly selected to initiate the contagion procedures and the procedures are repeated 600 times for each $p$.
Note that the slope of the linearity between $\alpha_c$ and $c$ is -2 in the most random design. The reasons are that, first, under the circumstance of a dense network, the market value of the failed stock is extremely small compared to the market value of all stocks one investors hold, so $\frac{\sum_{f \in F_\tau} w_{f,m,\tau=0}}{A_{m,\tau}}$ in Eq.~(\ref{eq:boundary}) is still small (recall that it is the $\tau=0$ that matters). Second, the scheme of randomly linking investor nodes and stock nodes makes the probabilities same for investors holding the failed stocks or not holding the failed stocks, i.e., $\frac{\sum_{m\in L_\tau} \alpha  w_{i,m,\tau}}{\sum_m w_{i,m,\tau=0}} =
\frac{\sum_{m \notin L_\tau} w_{i,m,\tau}}{\sum_m w_{i,m,\tau=0}}=0.5$ in Eq.~(\ref{eq:boundary_sim}), which gives $\alpha_c=1-2c$.} \label{fig:random}
\end{figure}

One of the possible ways to augment system resilience is to adjust the investing pattern on the whole to avoid the dominance of small but critical stocks. To achieve this, we conduct a series of straightforward random experiments, see Fig.~\ref{fig:random}.
When we completely randomize the original network, the linearity between $\alpha_c$ and $c$ has become steeper, i.e., $\alpha_c=1-2c$, which is good as the critical market confidence is lower at a certain level of price limit as compared to the original case. In fact, even with randomizing of a part of the original network, the linearity of $\alpha_c=1-c$ would be successfully adjusted. This implies that the system resilience could be improved by regulating the whole picture of investment pattern whereas retaining part of the original investment structure. 
In essence, the randomizations have successfully modify the distribution of both nestedness and branching (see Fig.~S10). The nodes with high nestedness are gradually eliminated with the increase of randomization, indicating the extent of portfolio overlapping has been reduced. As a result, it would be more difficult for the driving nodes to take over failures and the system resilience would then be promoted. 
Similarly, the number of nodes with high branching have been significantly curtailed, implying the risk broadcasting power and the exposure to risk of driving nodes will be lowered accordingly. The adaption of the exact relationship between $\alpha_c$ and $c$ through randomizing the network in Fig.~\ref{fig:random}, in turn, highlights the importance of nestedness and branching in terms of system resilience.
The exploratory experiments show that imposing the variated investment strategy on the whole or partly can thus enhance the resilience of the system. More broadly, if market participants could make a compromise between individual profit maximization and system stability enhancement, weakening the herding behavior and rethinking the over-diversified investment strategy simultaneously, the market structure will be reformed as the nestedness and branching are redistributed. As a result, the system resilience would be boosted and the price limit will work better for stabilizing the market. 

The unexpected side effect of the principle of profit maximization and risk minimization in the individual level of investors is inherently missed in existing understandings of portfolios. While our results both theoretically and empirically suggest that from the view of system resilience, overlapping portfolios due to herding investments derived from this principle unintentionally forge the emergence of small yet critical stocks that drive the market to collapse. 
In terms of networking investors and stocks, ideas from system science can help manifest the market crash and inject new insights to the practice of market supervision. And to obtain these insights might be challenging for the classical approaches in finance. Even more importantly, the crash of stock market also offers a new testbed to examine the previous understandings of system science and surprisingly, small nodes, which are conventionally thought to be trivial in risk contagion, emerge to be the most critical parts that reignite the failure cascading and determine the system behavior at the critical phase. 
By lowering the disassortativeness of the system, the entanglement between small but critical parts and instability of the whole system can be effectively weakened, thus leading to enhancement of system resilience. 
Our results may be of interest to policy markers tasked with developing regulation to promote market-wide stability and venue operators interested in designing effective trading strategies.

\section*{Methods}
\subsection*{Data}
The dataset contains the market value of shares held by mutual fund for listed stocks on June 30 2015, around the period that the severe stock market crash happened. It covers 1512 mutual funds and 2709 stocks listed on either Shanghai Stock Exchange or Shenzhen Stock Exchange. 
%The number of stocks listed in the A-share market are 2956 by then but some stocks has no investment from any mutual funds, and thus are excluded. The mutual funds are all those disclosed their holdings in China. 
\citet{lu2018herding} has pointed out that though the ownership data is taken at one particular time of the year, it represents noisy yet unbiased estimate of mutual funds’ investment preferences in that year or at least the days around the reporting date. 
We group ownership by mutual fund management companies as we assume that mutual funds under the same management company could have collective actions on an individual stock. The edge weights are equal to the sum of market value hold by mutual funds under the same management institutes. There are 87 mutual fund companies and 2709 stocks included in the study. Additionally, we deploy the proposed model into the mutual fund and stock network, in which the 1512 mutual funds are used as in the investor entities in Fig.~\ref{fig:example} instead of the 87 mutual fund companies. The network is less denser and the results could be found in Fig.~S8 and Fig.~S9, which are consistent with those from the network of mutual fund companies and stocks. 
%More discussion on the effectiveness of the dataset toward the study of 2015 Chinese crash could be found in~\citet{lu2018herding}.

The timing of stocks reaching their limit down prices are obtained by integrating two sources of information: the statuses of stocks and the stocks' intraday prices. The statuses of stocks acknowledge us whether the stocks reach their limit down prices or not during the trading day. If they did, we pick out the time that the stocks first reached their lowest prices of the day, i.e., their limit down prices, as their moments of failures. Data are downloaded from \textit{Wind Information}.

\subsection*{Contagion model}
We develop the following contagion model with market confidence $\alpha \in [0,1]$ and price limit $c \in (0,1)$. 

Step 1: $\tau=0$, we initially shock a single stock $i$, wiping out $c$ of its market value and it is then considered as failed. We have $S_{m,\tau=1}=(1-c)S_{m,\tau=0}$, $w_{i,m,\tau=1}=(1-c)w_{i,m,\tau=0}$.

Step 2: Updata the stocks' market value $\forall i$, $S_{i,\tau}=\sum_m w_{i,m,\tau}$. %At $\tau=0$, this equals to Step 1.

Step 3: If $\forall i$, $\frac{S_{i,\tau}-S_{i,\tau=0}}{S_{i,\tau=0}}>-c$, no further failures, the algorithm ends. If $\exists i$, $\frac{S_{i,\tau}-S_{i,\tau=0}}{S_{i,\tau=0}}\leq-c$, we call these stocks failed and add them into the stocks list $F_\tau$. The set of neighbors of stocks belong to $F_\tau$ is denoted as $L_\tau$.

Step 4: Delete the stocks in $F_\tau$ as they reach their down limit prices and are regarded as completely illiquid. Update the investors holding market value, $A_{m,\tau}=\sum_i w_{i,m,\tau}$.

Step 5: $\tau=\tau+1$, update the investors holding values, i.e., $\forall m \in L_\tau$, $\forall i$ , $w_{i,m,\tau+1}=\alpha \frac{A_{m,\tau+1}}{A_{m,\tau}}$. The term $\frac{A_{m,\tau+1}}{A_{m,\tau}}$ defines the degree of illiquidity results from the price limits of failed stocks in which we assume that the illiquidity of holding portfolio, a depreciating investor may depress the price of those stocks in the market. The confidence effects on stock prices depression are mild or negligible when $\alpha=1$, but become more severe as $\alpha$ decrease. 

Step 6: Return to Step 2.

\subsection*{Theoretical explanation: the nestedness and branching}\label{sec:analy}

By mapping our model onto a generalized process, we show analytically here that there is a region in parameter space where further cascades of failures occur. Denote the market value of investor $m$ holding stock $i$ as $w_{i,m,\tau}$, i.e., the edge weight from node $m$ to node $i$. The original market value of stock $i$ is $S_{i,\tau=0}=\sum_m w_{i,m,\tau=0}$. The original market value of investor is $A_{m,\tau=0}=\sum_i w_{i,m,\tau=0}$. Denote the stocks fail at $\tau$ as $F_\tau$. Denote the investors holding stocks that fail at $\tau$ as $L_\tau$. Considering the devaluation in Eq.(\ref{eq:devalue}), the stocks' failure boundary in Eq.(\ref{eq:down_limit}) could be written as
\begin{equation}\label{eq:boundary}
\frac{\sum_{m\in L_\tau} \alpha (1-\frac{\sum_{f \in F_\tau} w_{f,m,\tau=0}}{A_{m,\tau}}) w_{i,m,\tau} + \sum_{m \notin L_\tau} w_{i,m,\tau}}
{\sum_m w_{i,m,\tau=0}} \leq 1-c,
\end{equation}
where $m$ belongs to the investors stock $i$ connects. Following Eq.(\ref{eq:down_limit}) and (\ref{eq:devalue}), for every initially shocked stock, the market confidence needed to avoid its neighboring stock $i$'s failure at $\tau+1$ could be calculated as 
\begin{equation}\label{eq:ind_alpha_c}
\alpha_{c_i} =\frac{(1-c)\sum_m w_{i,m,\tau=0}-\sum_{m \notin L_\tau} w_{i,m,\tau=0}}
{\sum_{m\in L_\tau} (1-\frac{\sum_{i \in F_\tau} w_{i,m,\tau=0}}{A_{m,\tau}}) w_{i,m,\tau=0}}.
\end{equation}

Consider $\tau=0$, we assume that $\frac{\sum_{f \in F_\tau} w_{f,m,\tau=0}}{A_{m,\tau}}$ is rather small because the investors have a wide range of portfolios and one of them would not be comparable with the investors' total holding values. See Fig.~S3(d) for evidence in the Chinese case. Eq.(\ref{eq:boundary}) would then be simplified as 
\begin{equation}\label{eq:boundary_sim}
\frac{\sum_{m\in L_\tau} \alpha  w_{i,m,\tau} + \sum_{m \notin L_\tau} w_{i,m,\tau}}
{\sum_m w_{i,m,\tau=0}} \leq 1-c.
\end{equation}
The intuition behind the numerator is that the market value of stock $i$ consists of two parts: one part held by investors connect to the failed stocks, the other part held by investors do not connect to the failed stocks. Apparently, the ratios of the two are critical in the network cascading. Inspired by this, two indicators for stocks are raised: nestedness and branching. 

The nestedness of stock $i$ on stock $j$ is the degree of how stock $i$ would be influenced by stock $j$'s failure, which could be defined as 
\begin{equation}
\text{nestedness of stock }i \text{ on stock }j=\frac{\text{the number of common neighbors between } i \text{ and } j}{\text{the degree of }i}.
\end{equation}
Nestedness basically measures the severity of portfolio overlapping. Note that the nestedness of stock $i$ on stock $j$ is not equal to the nestedness of stock $j$ on stock $i$. For instance, suppose stock $i$ is held by few investors while these investors hold the other stock $j$, then the nestedness of stock $i$ on stock $j$ would be one (see Fig.~\ref{fig:example}(c)). But if the stock $j$ has more investors, the nestedness of stock $j$ on stock $i$ would be small.
A higher value of nestedness implies a higher value of $\frac{\sum_{m \in L_\tau} w_{i,m,\tau}}{\sum_m w_{i,m,\tau=0}}$ in Eq.(\ref{eq:boundary_sim}) if edge weights are really close to each other, which means the stocks with higher nestedness have greater potential to get infected by other stocks' failures and their failures will drive further rounds of risk contagion (see Fig.~\ref{fig:nestedness_branching}). 

The branching for stock $i$ is defined as 
\begin{equation}
branching = \frac{\text{the highest degree of } i\text{'s neighbors}}{\text{the degree of } i}.
\end{equation}
Branching takes account of the number of neighbors stock $i$ have by definition and thus reveals the probability of stock $i$'s exposure to other stocks' failures. Remember in each step of contagion, it is only the stocks which share common neighbors with failed stocks that are taken account into the devaluation and have the potential to reach price limits, i.e., Eq.(\ref{eq:boundary}).
Therefore, branching unfolds the overall extent of stock $i$'s exposure to random shocks. Additionally, branching also depicts the capacity of spreading risk because it correlates to the number of stocks that one failed stocks could pass the depression to others through common neighbors (see Fig.~\ref{fig:example}(d)). A high level of branching is basically an outcome of investors' highly diversified portfolios. Nestedness and branching altogether govern the potential for the spread of shocks through the network and it is shown that they help provide effective information into the causes of the potential dynamical behavior as well as influence system resilience.

\linespread{1}
%\section*{Reference}
%\bibliographystyle{abbrvnat}
%\bibliography{cascade1}

\vspace{-0.4cm}
\section*{Acknowledgments}
\vspace{-0.4cm}
This research was financially supported by National Natural Science Foundation of China (Grant Nos. 71420107025 and 71871006).

\section*{Additional information}
\subsection*{Data availability statement} The datasets analyzed during the current study are available in the figshare.com repository, https://doi.org/10.6084/m9.figshare.8216582.v2.

\newpage
\begin{appendices}  

\setcounter{figure}{0}   
\renewcommand\thefigure{S\arabic{figure}}   

\section*{Supplementary Figures}

\begin{figure}[h!]
	\begin{minipage}{0.5\linewidth}
		\centering
		\includegraphics[width =0.9\linewidth]{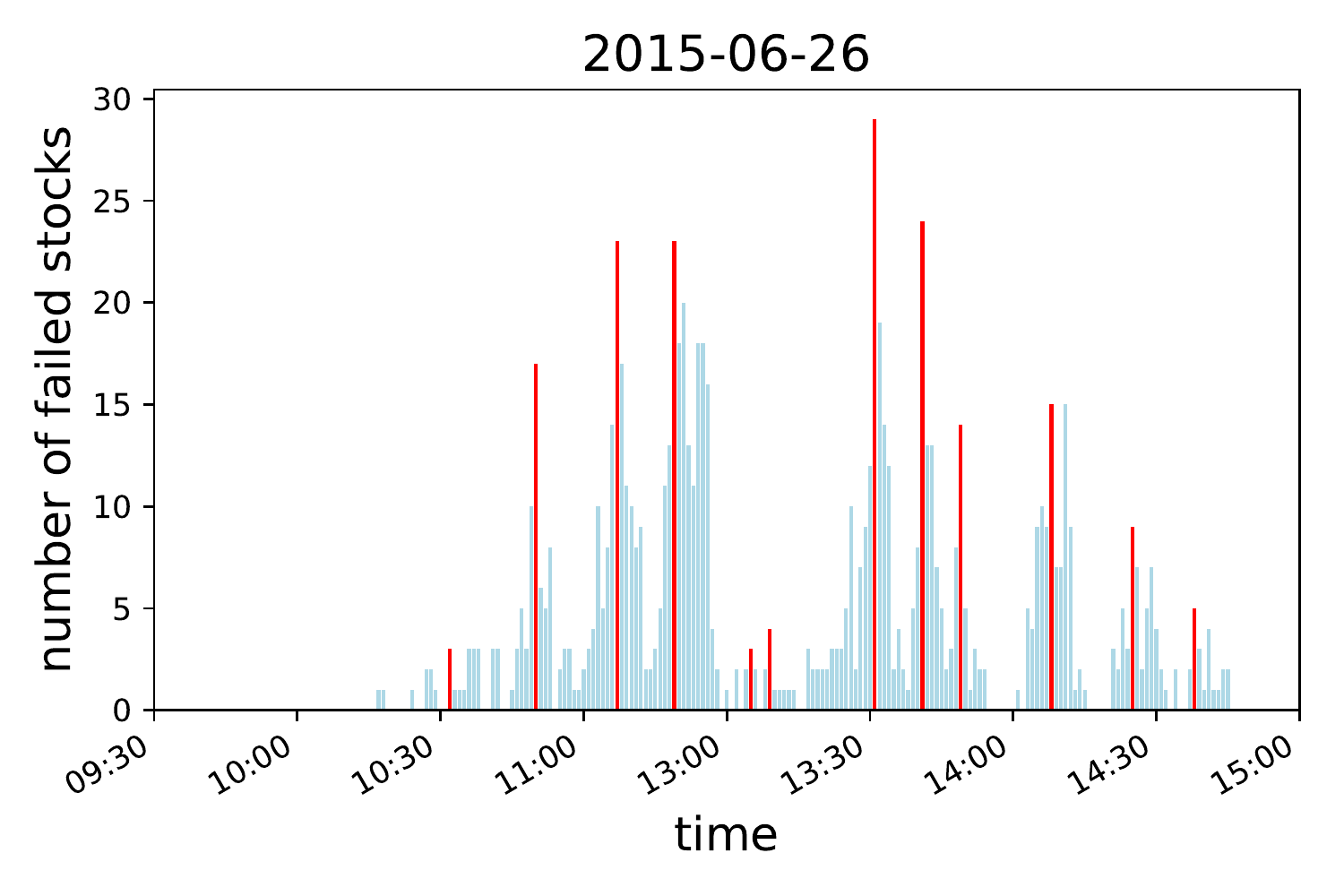}
	\end{minipage}
	\begin{minipage}{0.5\linewidth}
		\centering
		\includegraphics[width =0.9\linewidth]{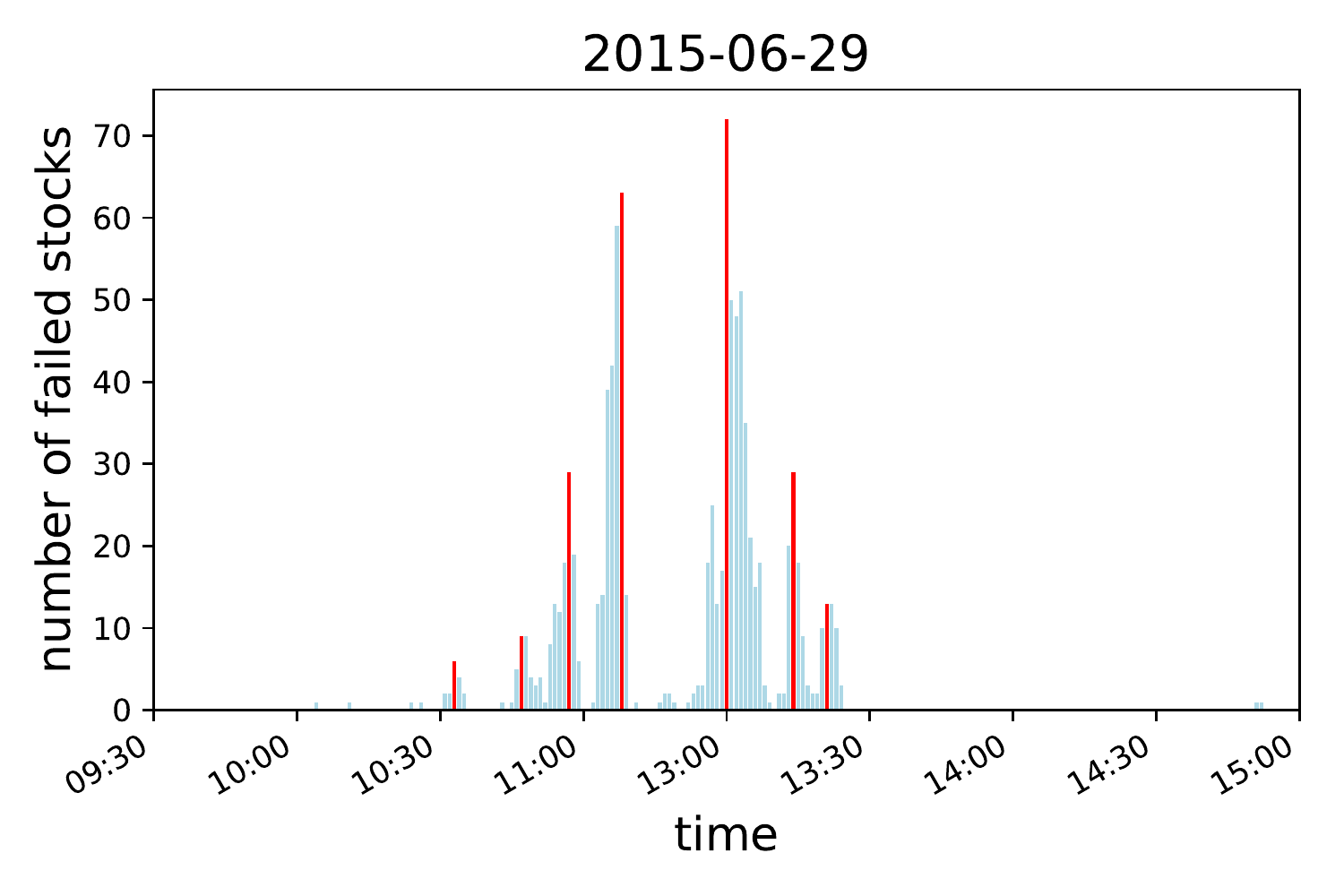}
	\end{minipage}

	\begin{minipage}{0.5\linewidth}
		\centering
		\includegraphics[width =0.9\linewidth]{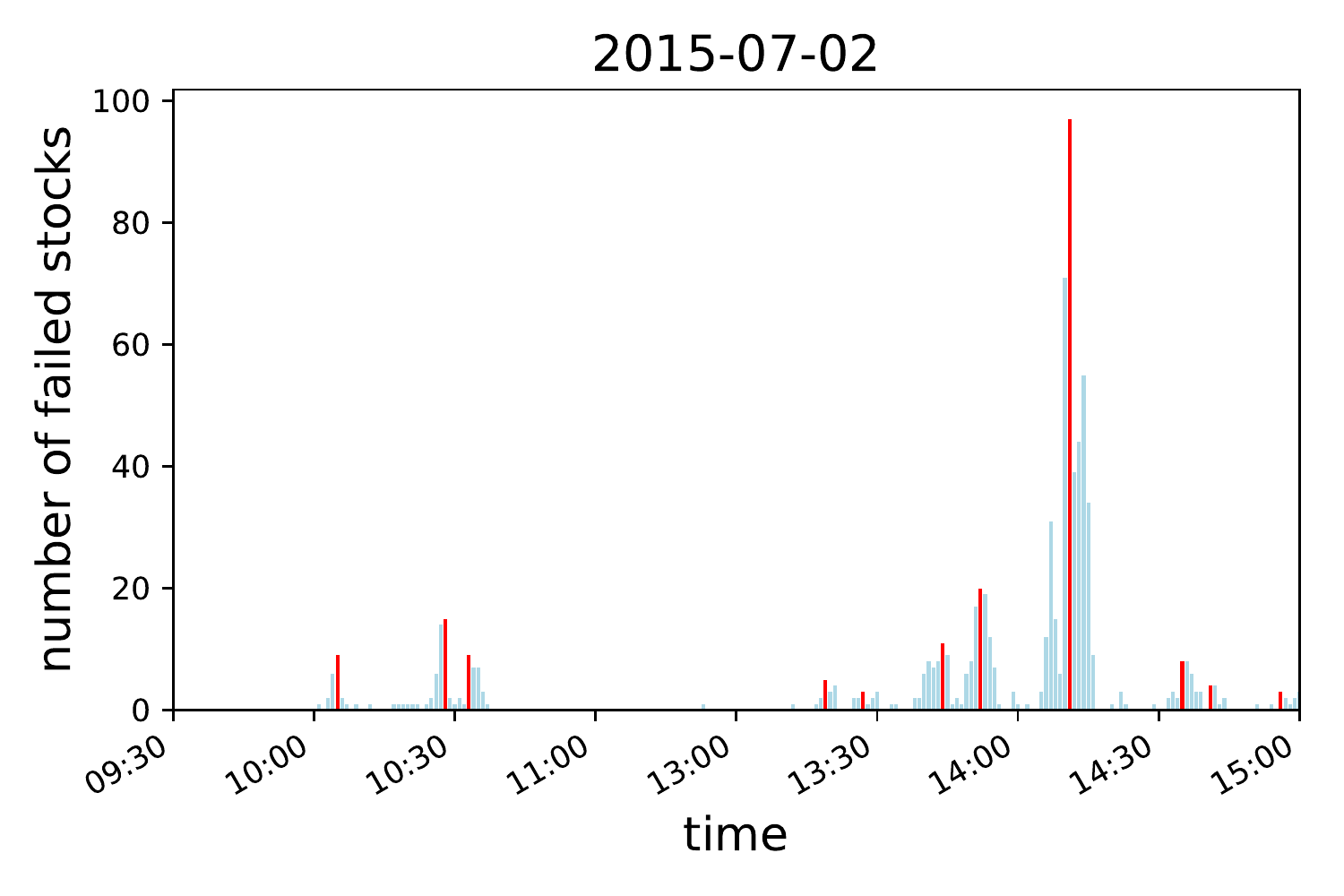}
	\end{minipage}
	\begin{minipage}{0.5\linewidth}
		\centering
		\includegraphics[width =0.9\linewidth]{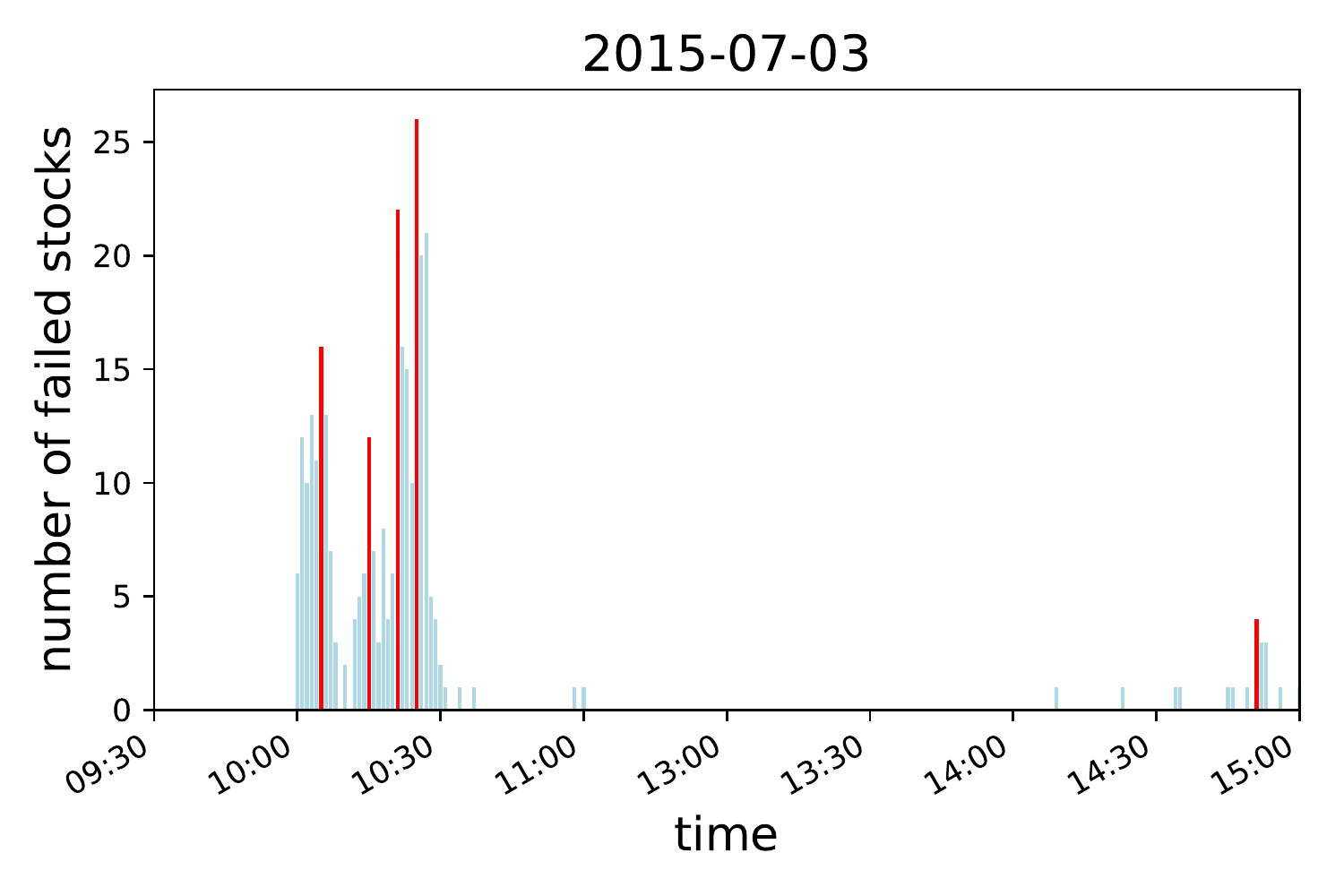}
	\end{minipage}
\caption{{\bf The waves of stocks reaching limit down prices during market crash.} The x-axis denotes the trading hours in minute-granularity from 9:30am to 11:30am and 13:00pm to 15:00pm. The y-axis is the number of stocks that reach their limit down prices, i.e., stocks that failed. The lunch breaks that range from 11:30am to 13:00pm everyday are neglected. Four trading days are regarded as market crashing days. A wave is defined as a time interval in which at least one stock reached limit down prices substantially in minute-granularity. The red bars denote the moments when the number of failed stocks reach the peak in a wave.}
\label{fig:four_days_peak}
\end{figure}

\newpage
\begin{figure}[h!]
\centering
\includegraphics[width =0.9\linewidth]{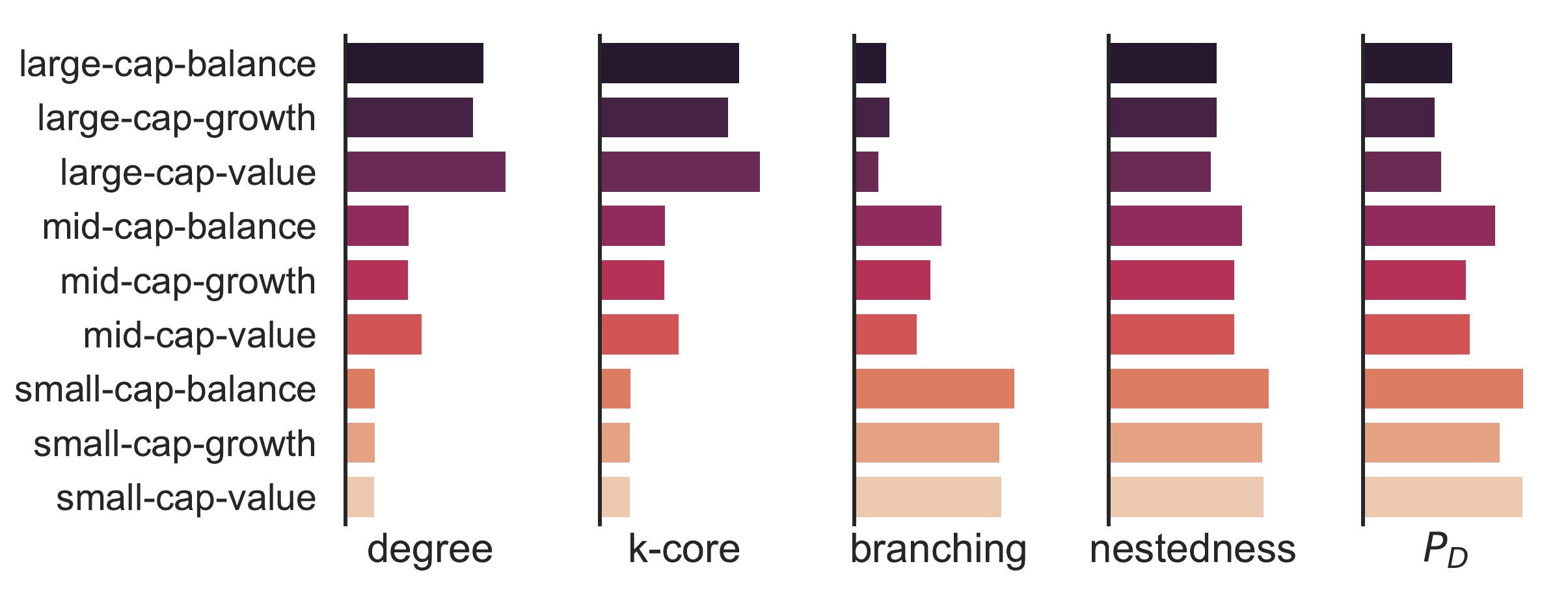}
\caption{{\bf The characteristic of different stocks.} The stocks are classified into nine groups according to market value and book-to-earning ratio. The classification is provided by \textit{Wind Information.} The x-axis ticks of the five indicators are omitted for clarity purpose.} \label{fig:cap}
\end{figure}

\newpage
\begin{figure}[h!]
	\begin{minipage}{0.5\linewidth}
		\centering
		{\footnotesize (a)}\\
		\includegraphics[width =0.9\linewidth]{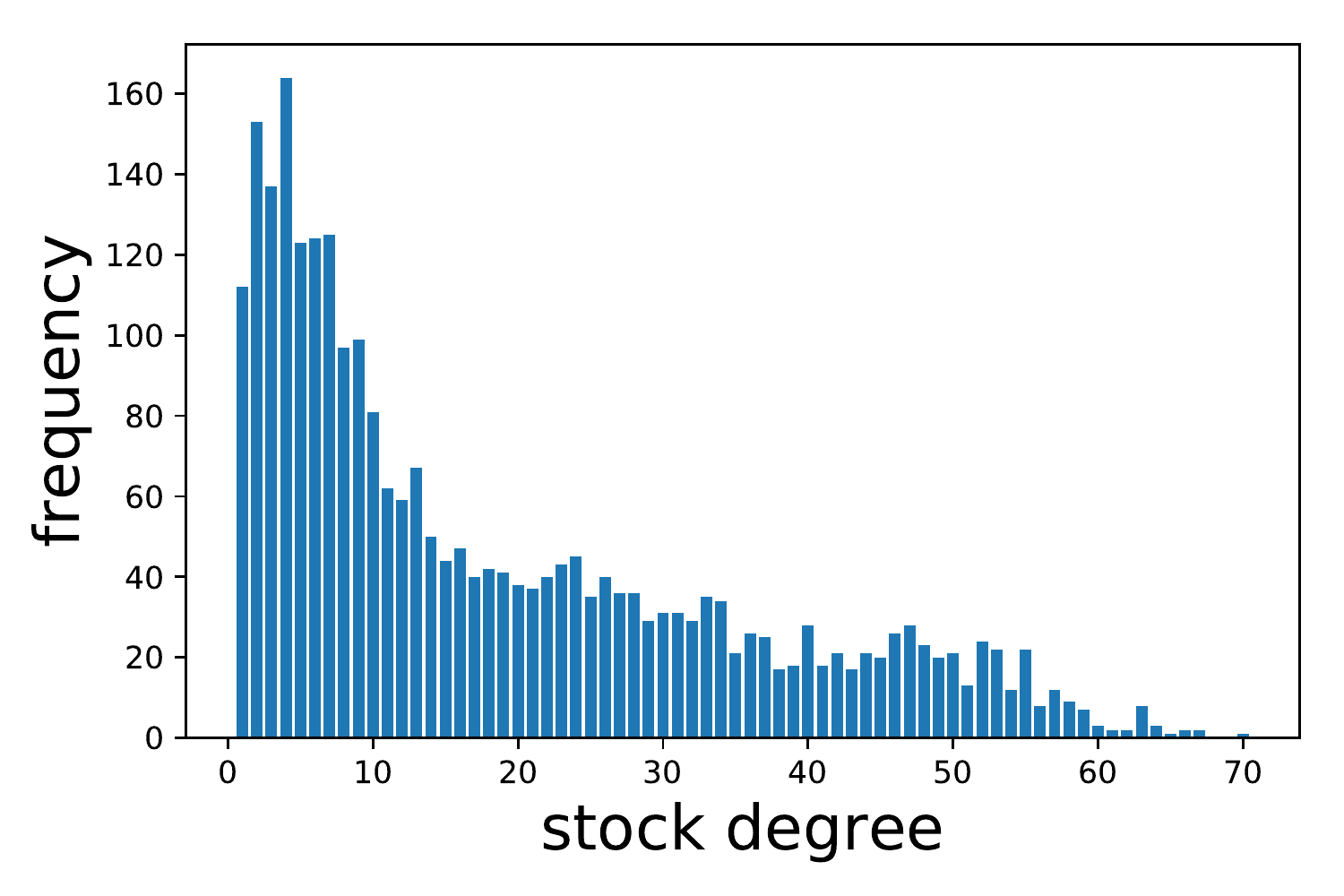}
	\end{minipage}
	\begin{minipage}{0.5\linewidth}
		\centering
		{\footnotesize (b)}\\
		\includegraphics[width =0.9\linewidth]{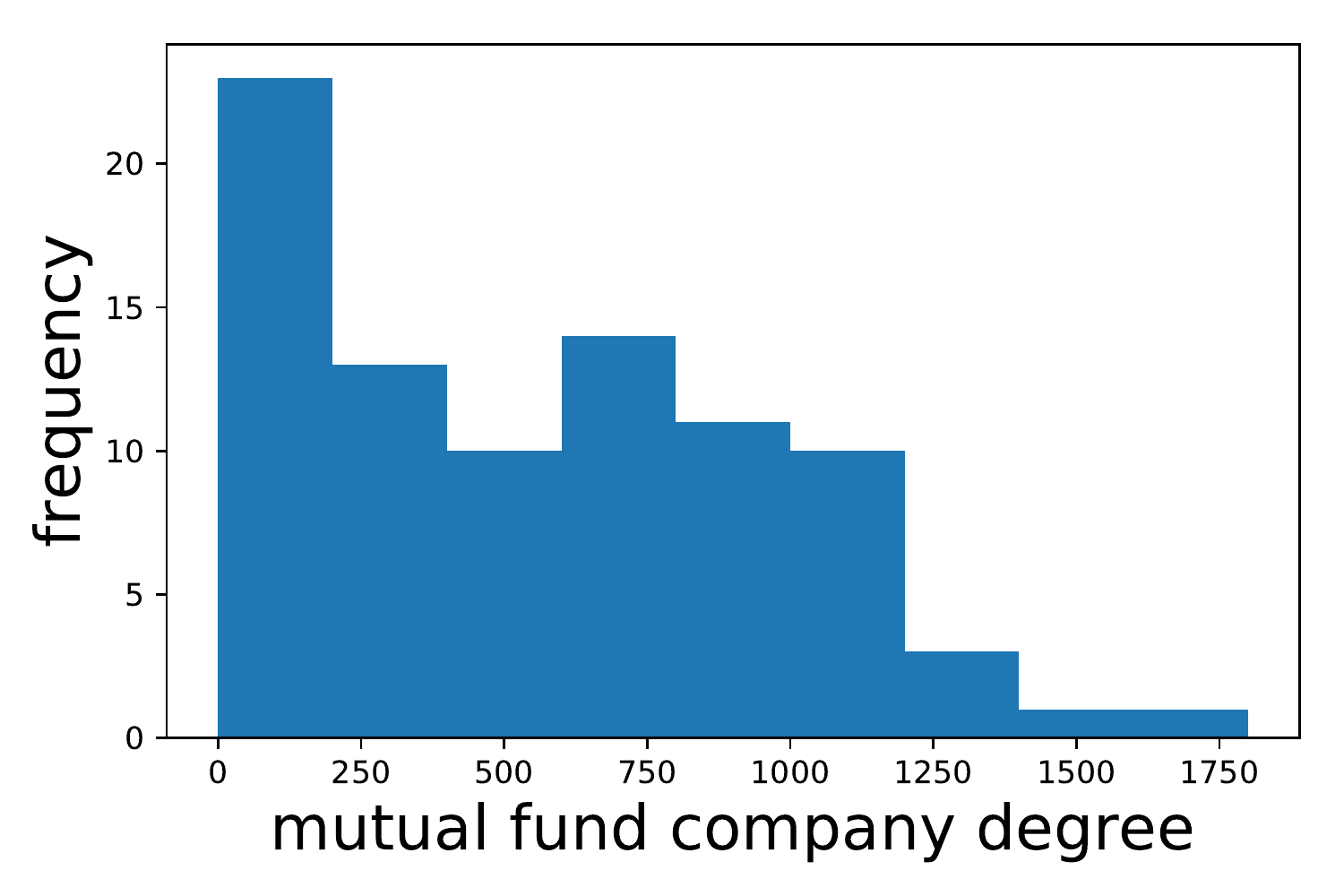}
	\end{minipage}
	\medskip
	
	\begin{minipage}{0.5\linewidth}
		\centering
		{\footnotesize (c)}\\
		\includegraphics[width =0.9\linewidth]{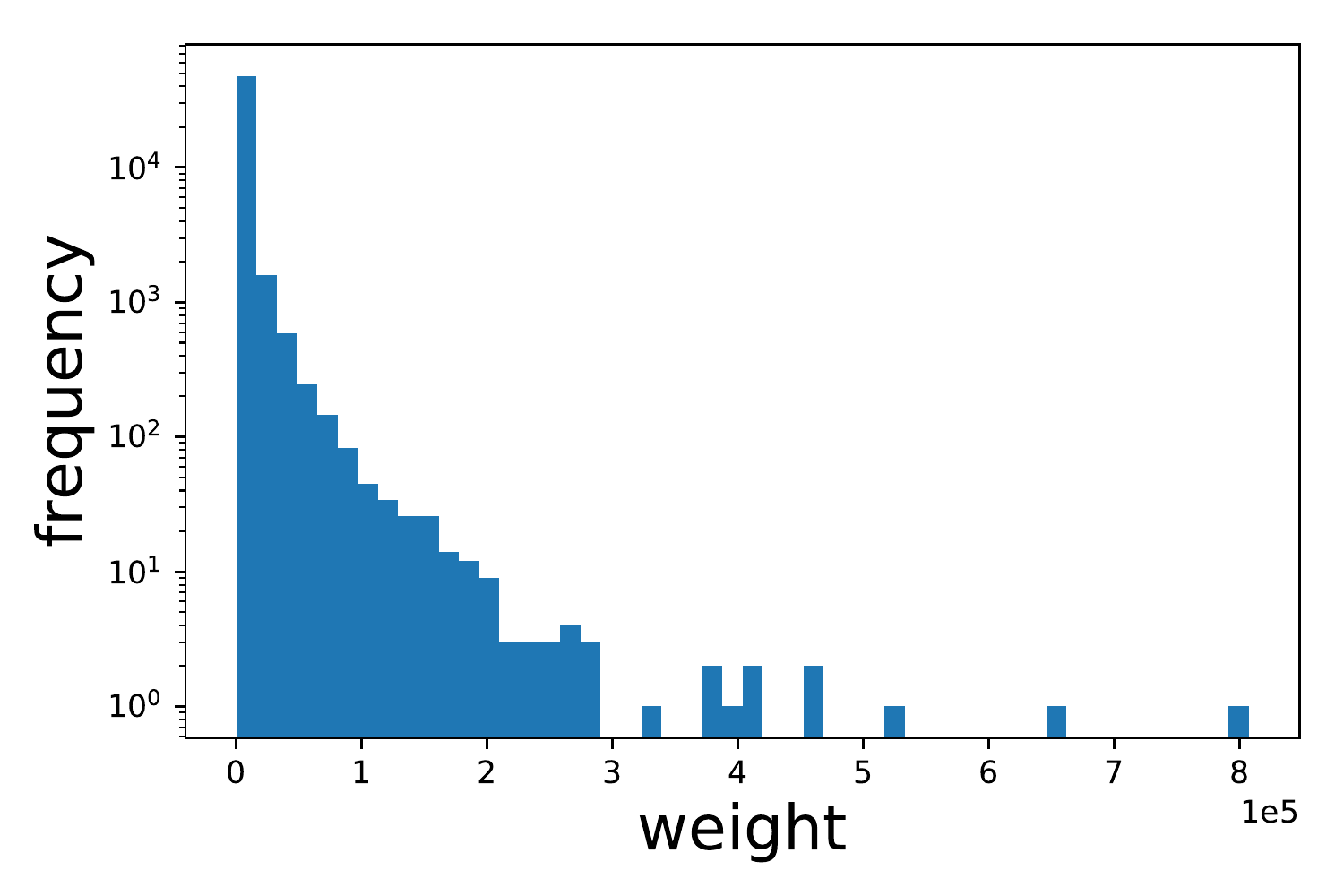}
	\end{minipage}
	\begin{minipage}{0.5\linewidth}
		\centering
		{\footnotesize (d)}\\
		\includegraphics[width =0.9\linewidth]{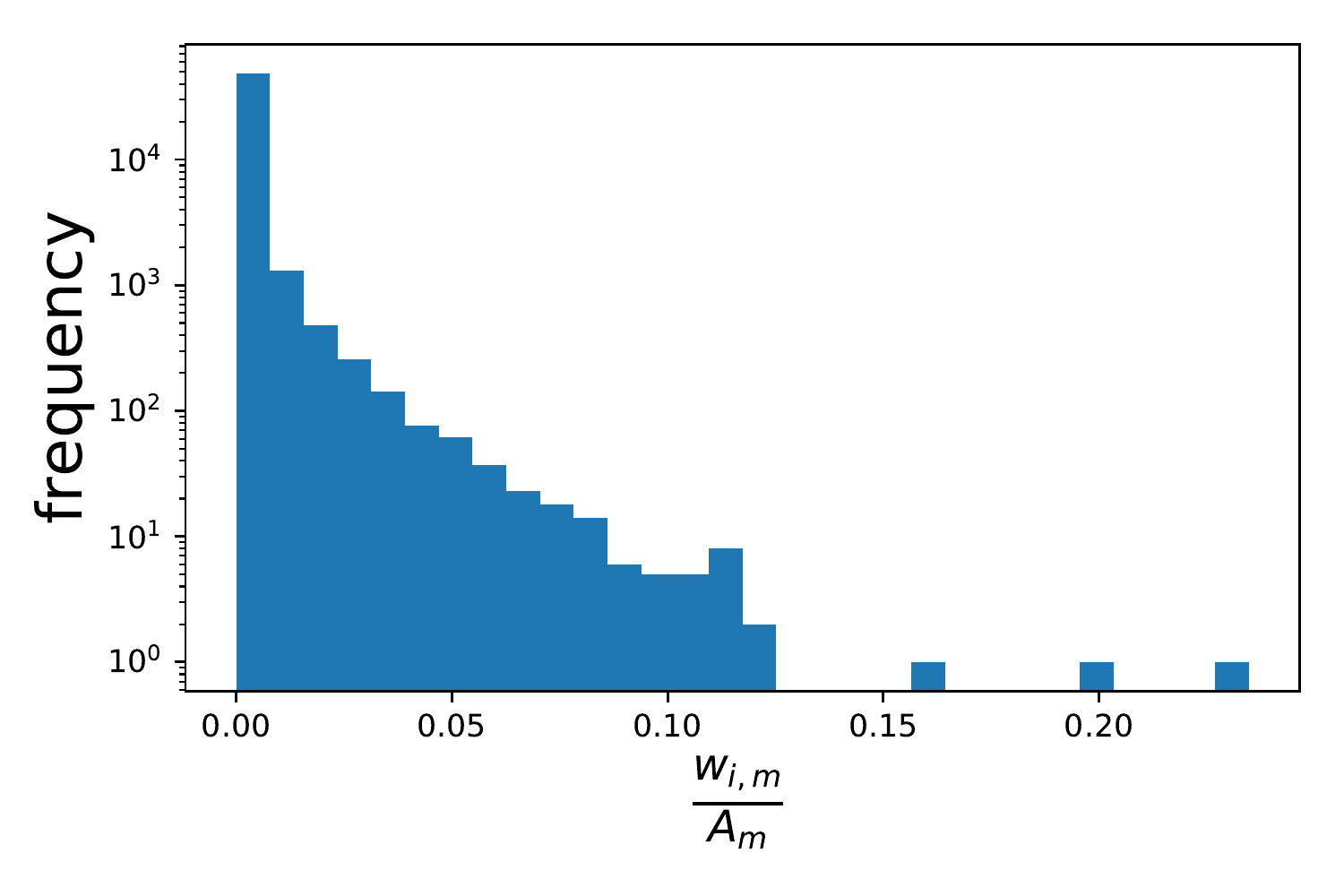}
	\end{minipage}
	
\caption{{\bf The attributes of mutual fund companies and stocks network.} (a) Degree distribution of stocks. (b) Degree distribution of mutual fund companies. (c) Weight distribution. (d) The weight ratios. $w_{i,m}$ is the market value of mutual fund company $m$ invests on stock $i$. $A_m$ is the stocks' market value that company $m$ holds. } 	
\label{fig:degree_distribution}
\end{figure}

\newpage
\begin{figure}[h!]
	\begin{minipage}{0.5\linewidth}
		\centering
		{\footnotesize (a)}\\
		\includegraphics[width =\linewidth]{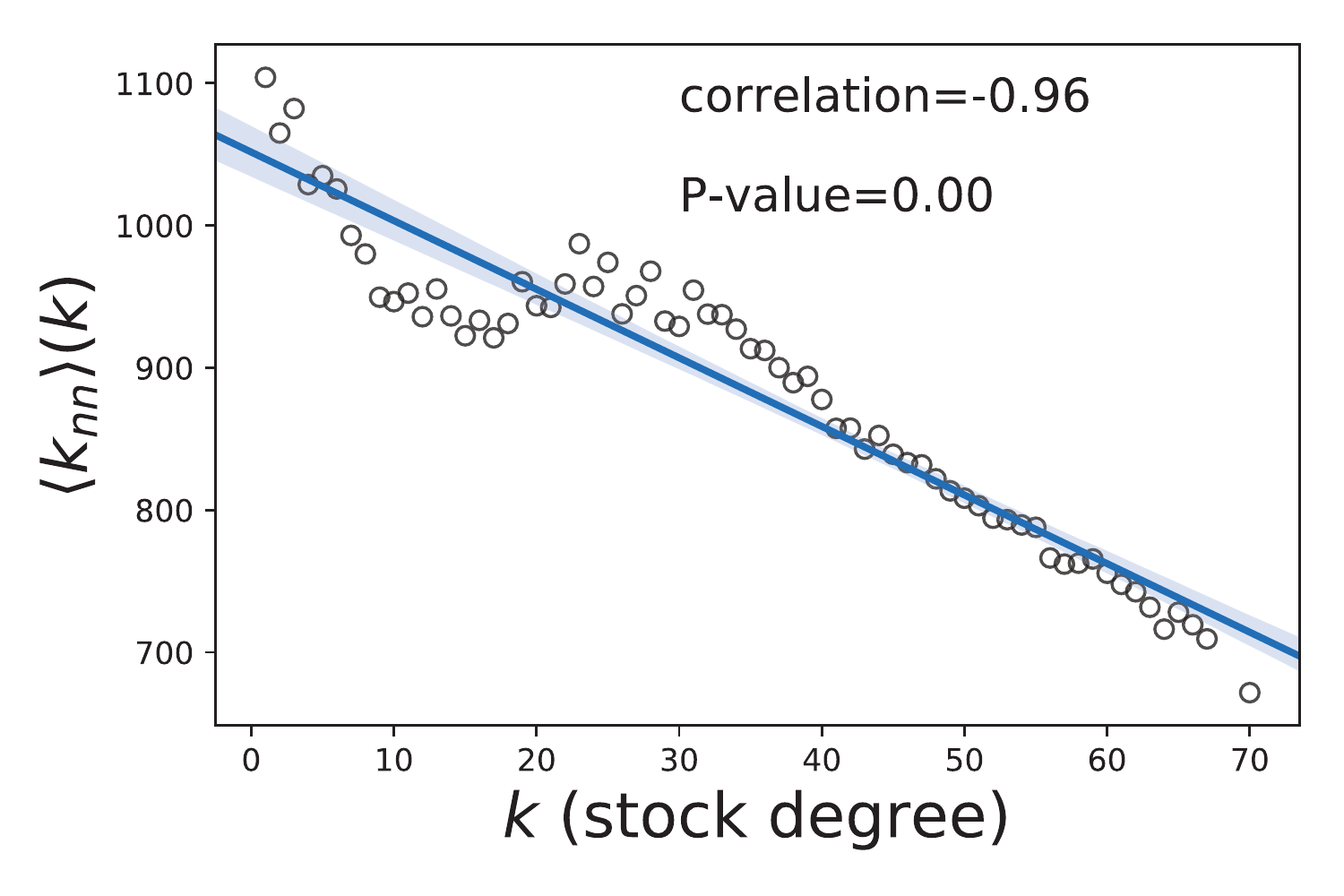}
	\end{minipage}
	\begin{minipage}{0.5\linewidth}
		\centering
		{\footnotesize (b)}\\
		\includegraphics[width =\linewidth]{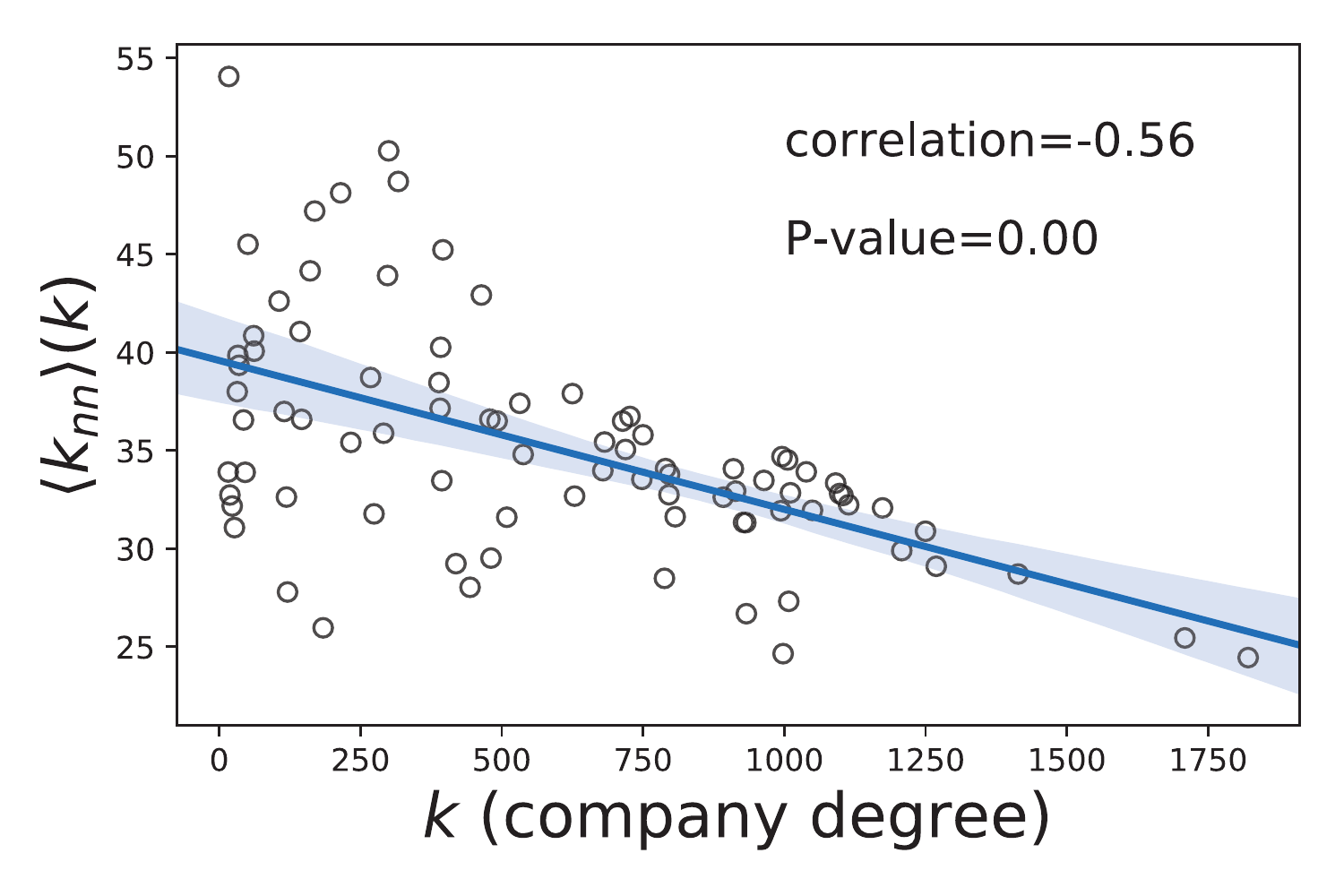}
	\end{minipage}
\caption{{\bf Degree disassortative.} $\langle k_{nn}\rangle (k)$ denotes the average nearest neighbor degree of nodes with degree $k$.
(a) X-axis is the degrees of stocks. The smaller the degrees of the stocks, the larger degrees of companies invest on them. (b) X-axis is the degrees of companies. The smaller the degrees of the companies, the larger degrees of stocks they would invest on. Overall, these two plots show that those who invest on the stocks with lower degrees have over-diversified portfolio strategies, and those who have mid-level diversified strategy would prefer the popular stocks (with high degrees). In addition, the companies holding stocks that enjoy popularity (with high degrees), though possess lower degrees on average, are still with degrees higher than 600, revealing the great potential for risk contagion. }\label{fig:excess}
\end{figure}

\newpage
\begin{figure}[h!]
	\begin{minipage}{0.5\linewidth}
		\centering 
		{\footnotesize (a)}\\
		\includegraphics[width =0.9\linewidth]{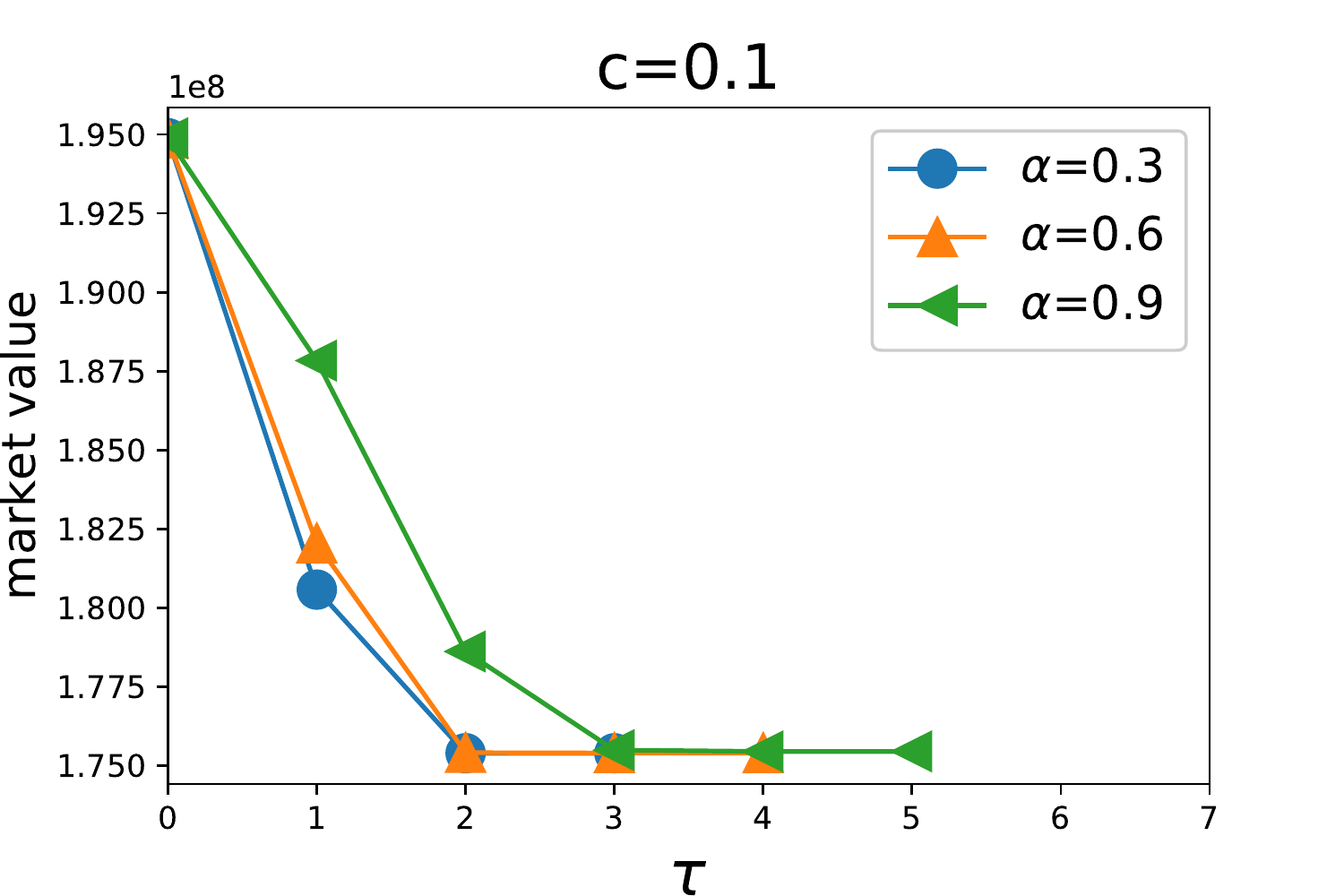}
	\end{minipage}\hfill
	\begin{minipage}{0.5\linewidth}
		\centering 
		{\footnotesize (b)}\\
		\includegraphics[width =0.9\linewidth]{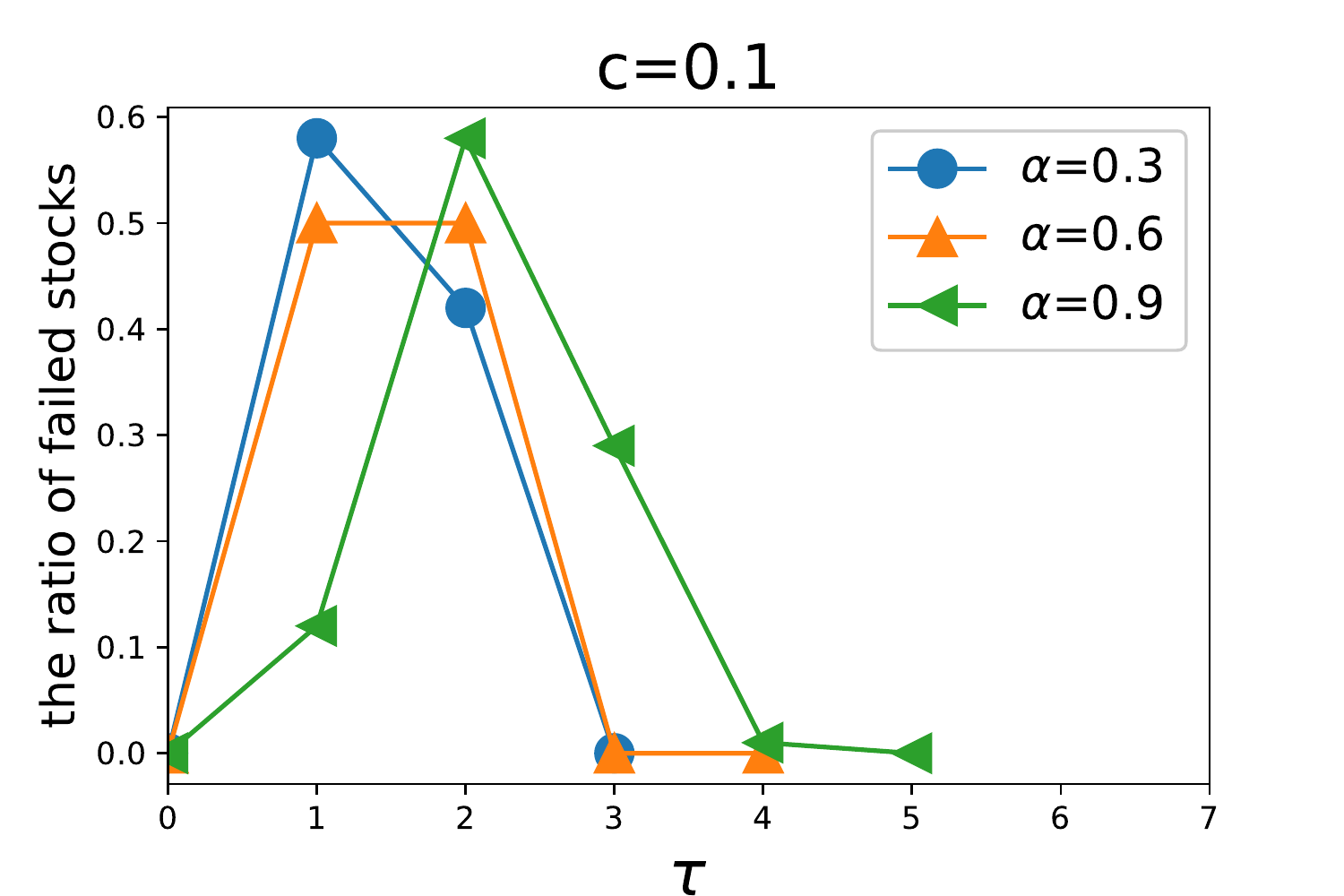}
	\end{minipage}
\caption{{\bf The system status for each step $\tau$ when $c=0.1$.} (a) The total stock market value left in the network in process of cascading failure. (b) The ratio of newly failed stocks in the corresponding time step. A high $\alpha$ can slow down the collapse of the system.} \label{fig:each_tau}
\end{figure}

\newpage
\begin{figure}[h!]
	\begin{minipage}{0.5\linewidth}
		\centering
		\includegraphics[width =\linewidth]{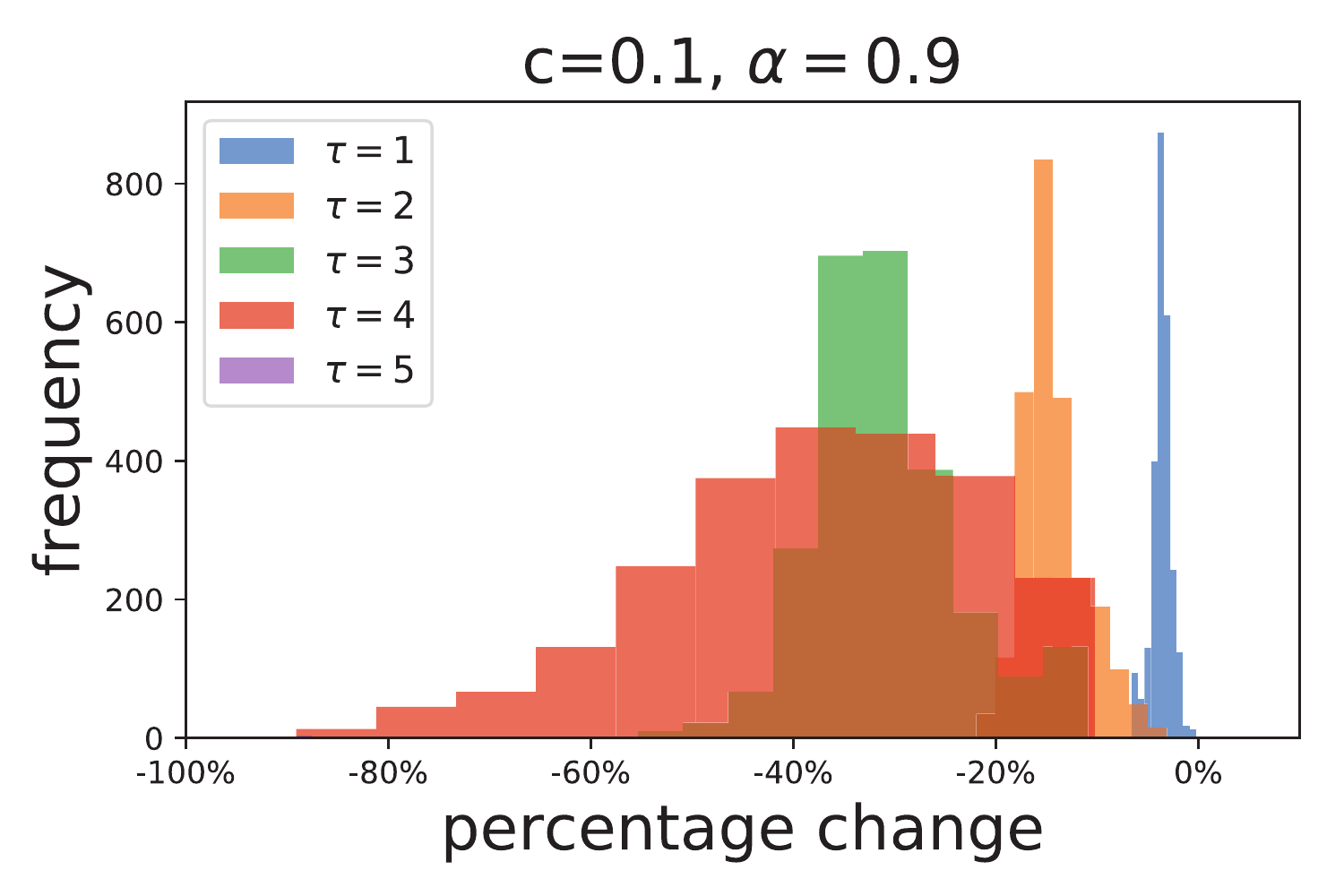}
	\end{minipage}
	\begin{minipage}{0.5\linewidth}
		\centering
		\includegraphics[width =\linewidth]{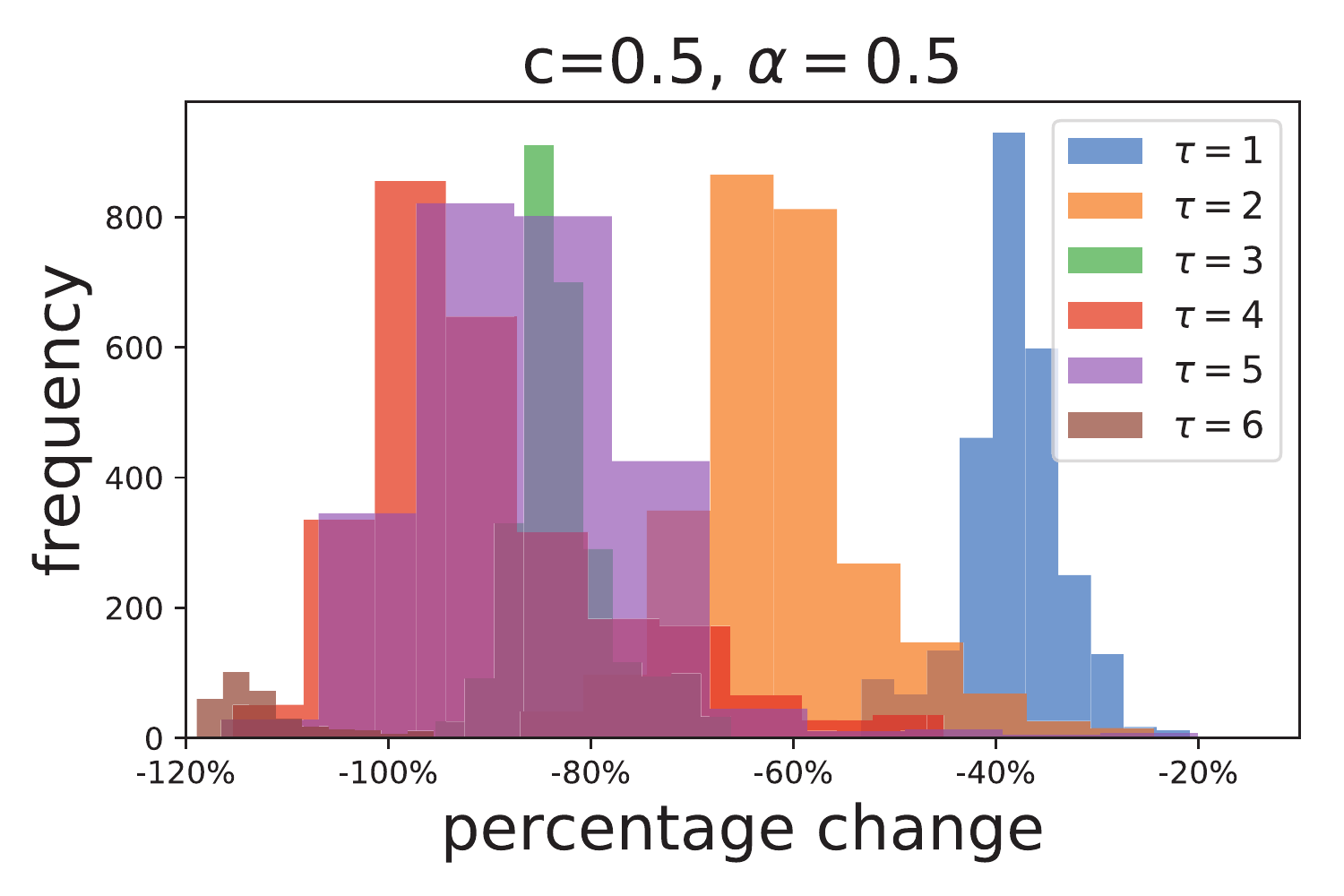}
	\end{minipage}
	\begin{minipage}{\linewidth}
		\centering
		\includegraphics[width =0.5\linewidth]{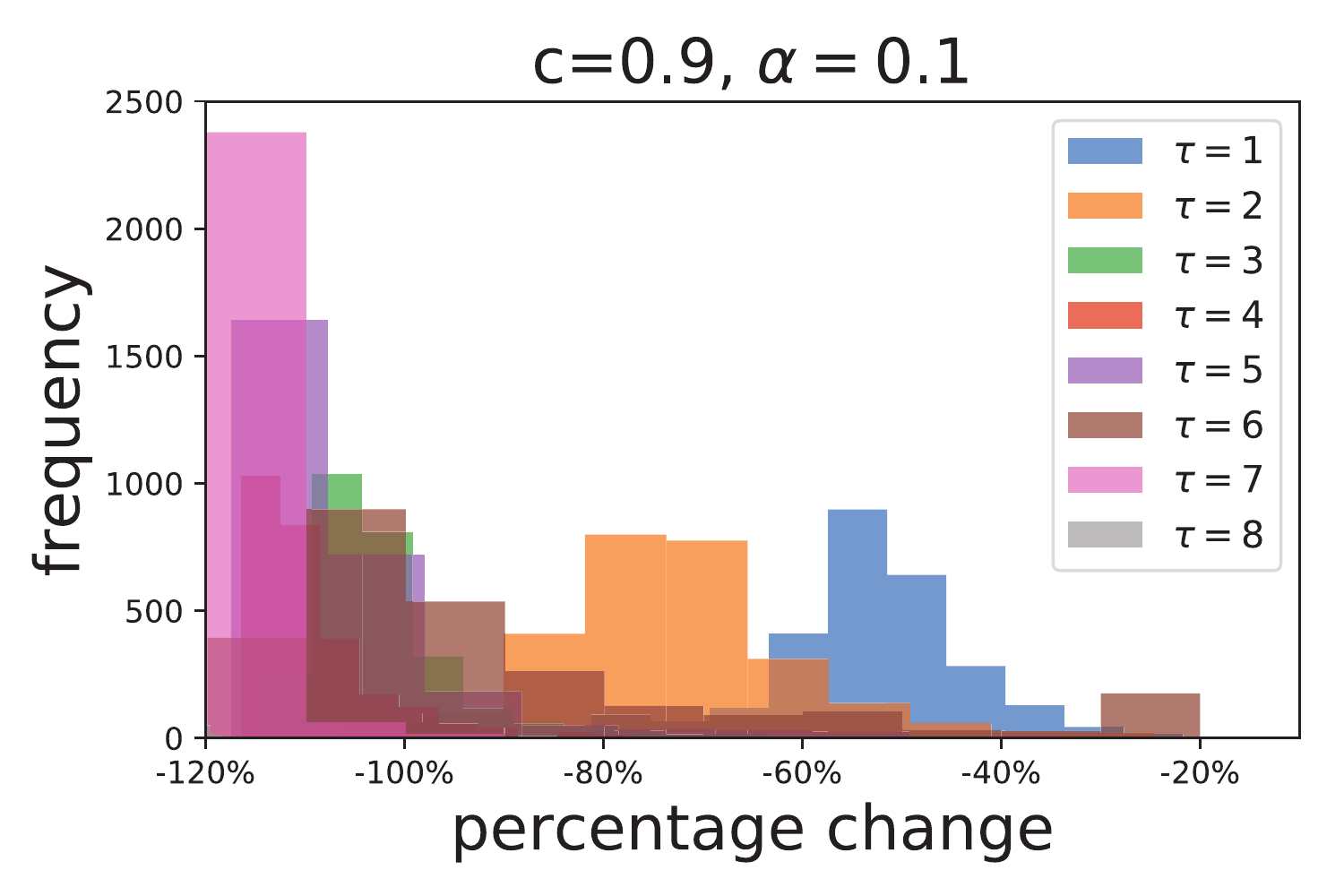}
	\end{minipage}
\caption{{\bf The distribution of active stocks' percentage changes at each time step $\tau$.} Stocks failed before $\tau$ step have been excluded in the network and the percentage changes of all the active stocks left are thus calculated. The stocks of percentage changes smaller than $-c$ will be the newly failed stocks at $\tau$. As can be seen, most failures happen before the final step. }\label{fig:percentage_change_tau}
\end{figure}

\newpage
\begin{figure}[h!]
\centering
\includegraphics[width= 0.8\linewidth]{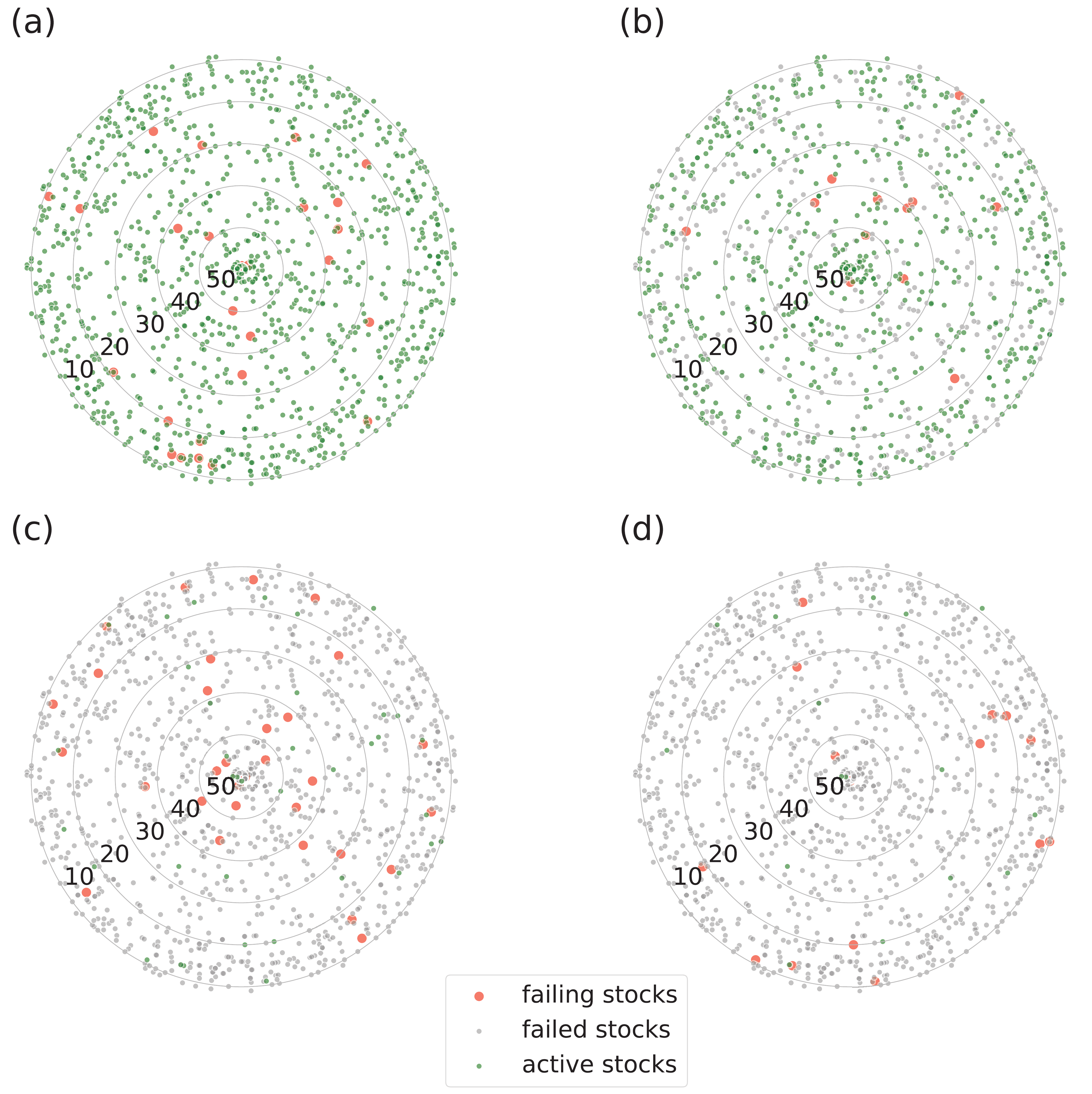}
\caption{{\bf Four snapshots on the cascading of stocks reaching the limit down prices on June 29, 2015 in Chinese stock market.} (a) Time interval is 9:25am~9:30am. (b) Time interval is 10:00am-10:30am. (c) Time interval is 13:30pm-13:30pm. (d) Time interval is 14:30pm-15:00pm. Failing stocks (red) are those reach down limit prices in the corresponding time interval. Failed stocks (grey) are those have already reached down limit prices before the starting point of the corresponding time interval. Active stocks (green) are those that have not reached down limit prices. The numbers in the graphs denote the $k$-core indexes. That is to say, the closer the nodes are to the core, the higher their $k$-core indexes are. It can be seen that the failures begin at the periphery, then move to the core, spread to the whole system and finally come back to periphery again.} %\normalfont{From (a) to (d) is from morning to afternoon.}
\label{fig:real_dynamic}
\end{figure}

\newpage
\begin{figure}[h!]
	\begin{minipage}{0.5\linewidth}
		\centering
		{\footnotesize (a)}\\
		\includegraphics[width =\linewidth]{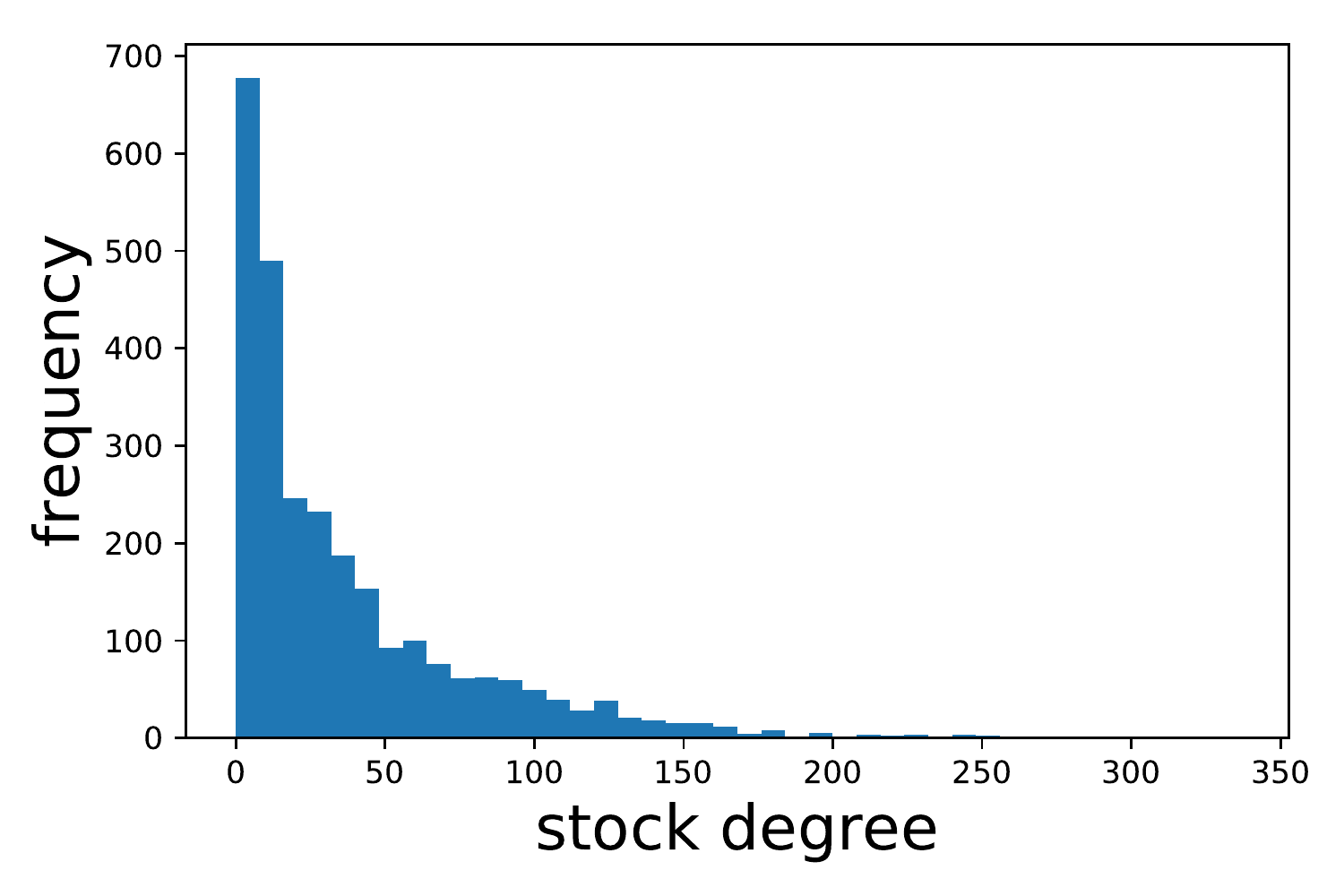}
	\end{minipage}	
	\begin{minipage}{0.5\linewidth}
		\centering
		{\footnotesize (b)}\\
		\includegraphics[width =\linewidth]{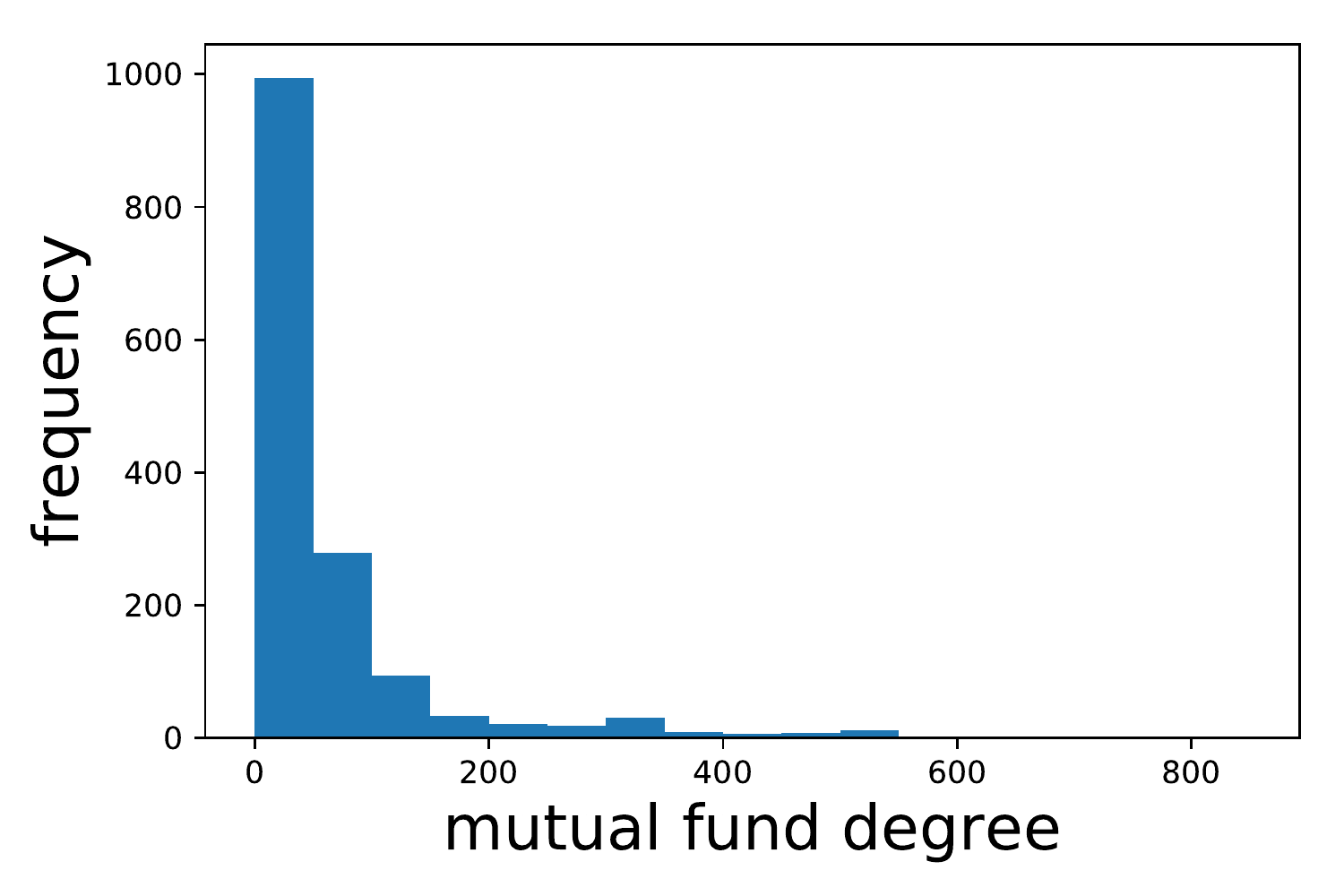}
	\end{minipage}
	
	\medskip
	\begin{minipage}{0.5\linewidth}
		\centering
		{\footnotesize (c)}\\
		\includegraphics[width =\linewidth]{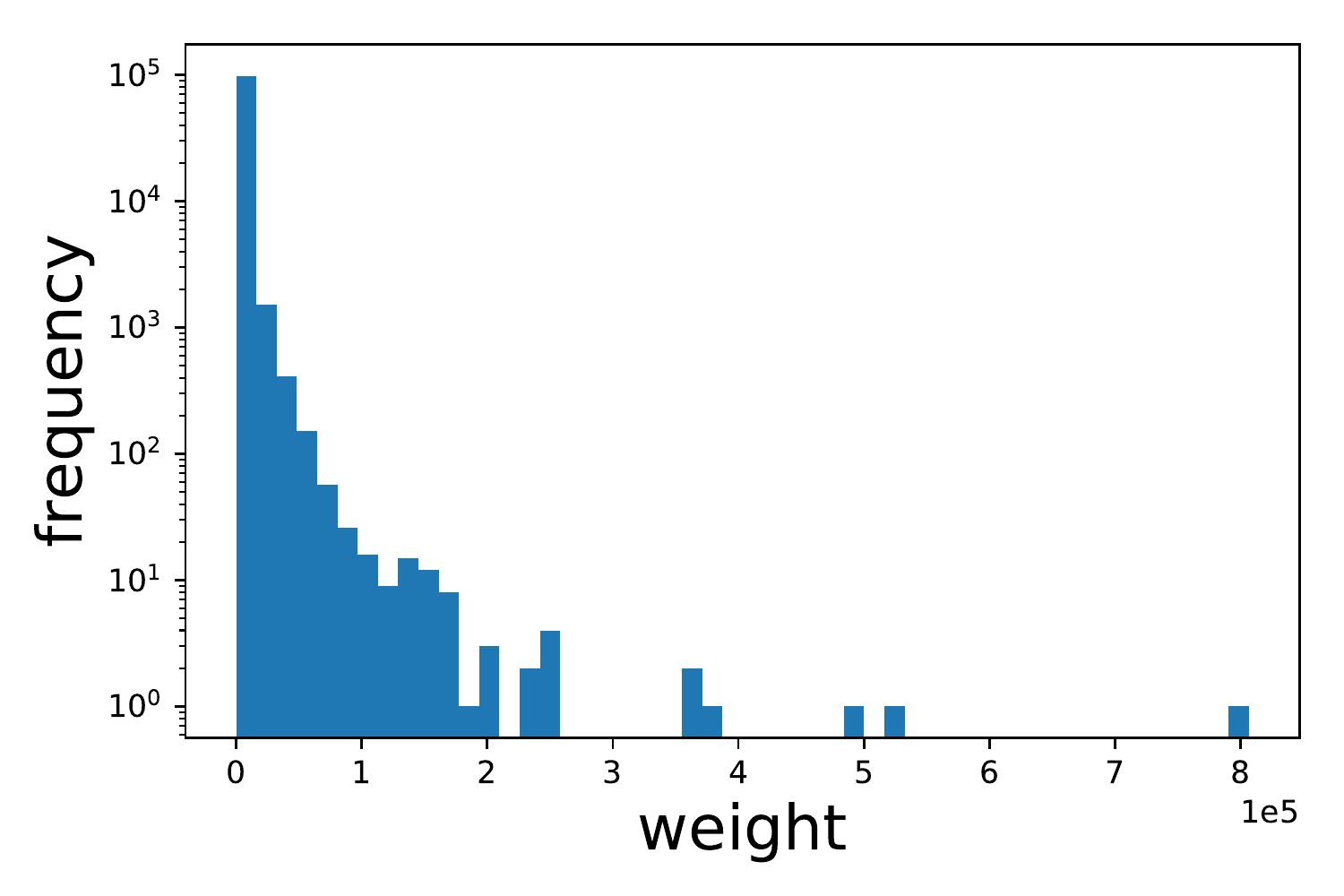}
	\end{minipage}
%	\begin{minipage}{0.5\linewidth}
%		\centering
%		{\footnotesize (d)}\\
%		\includegraphics[width =\linewidth]{fund_weight_power_law}
%	\end{minipage}
%	
%	\medskip
	\begin{minipage}{0.5\linewidth}
		\centering
		{\footnotesize (d)}\\
		\includegraphics[width =\linewidth]{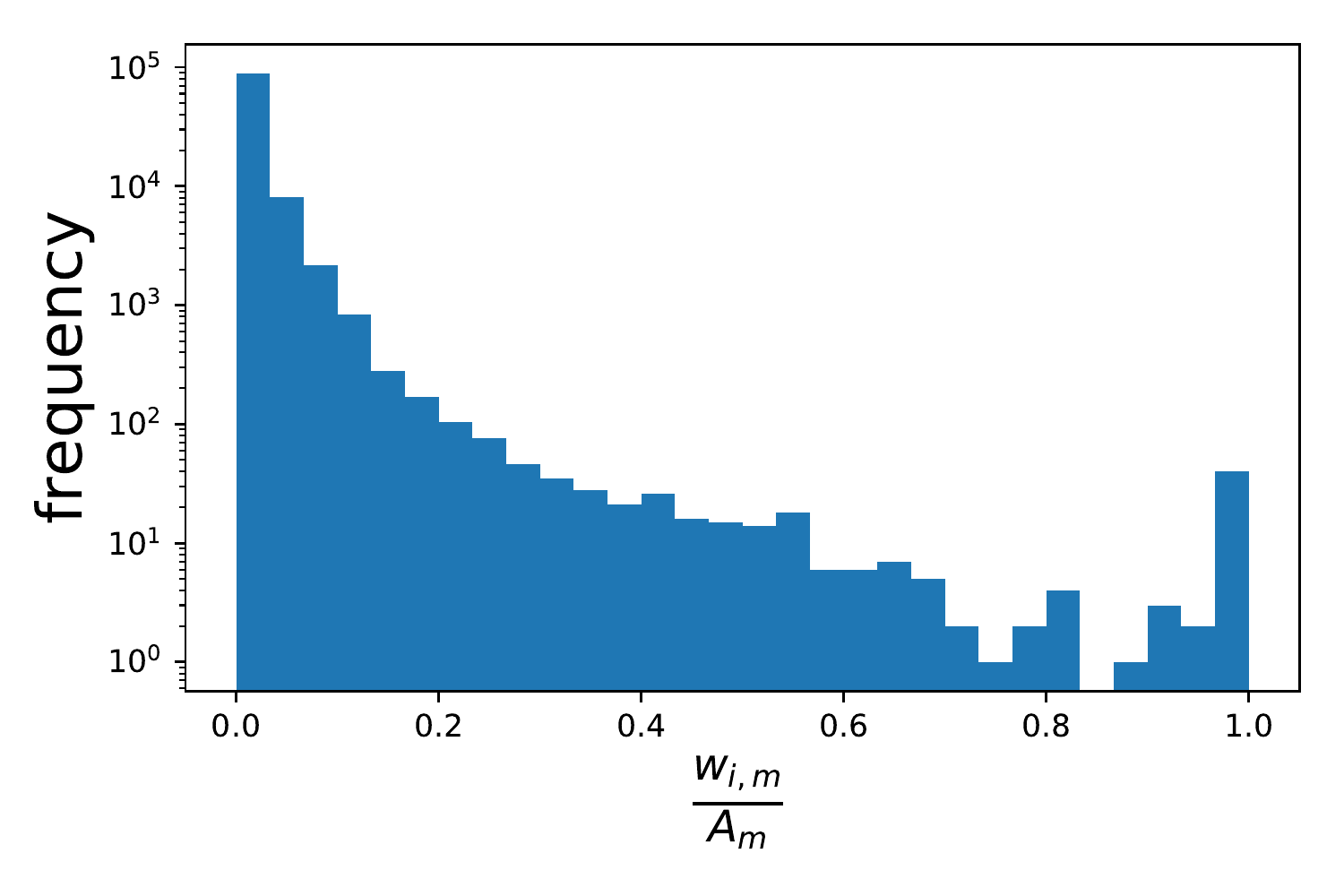}
	\end{minipage}
\caption{{\bf Mutual fund and stock network attributes.} (a) Degree distribution of stocks. (b) Degree distributions of mutual funds. (c) Weight distribution. (d) The weight ratios. $w_{i,m}$ is the market value of mutual fund $m$ invests on stock $i$. $A_m$ is the stocks' market value that fund $m$ holds. }\label{fig:mutual_fund_degree_distribution}
\end{figure}

\newpage
\begin{figure}[h!]
	\begin{minipage}{0.5\linewidth}
		\centering
		{\footnotesize (a)}\\
		\includegraphics[width =0.9\linewidth]{alpha_c.pdf}
	\end{minipage}
	\begin{minipage}{0.5\linewidth}
		\centering
		{\footnotesize (b)}\\
		\includegraphics[width =0.9\linewidth]{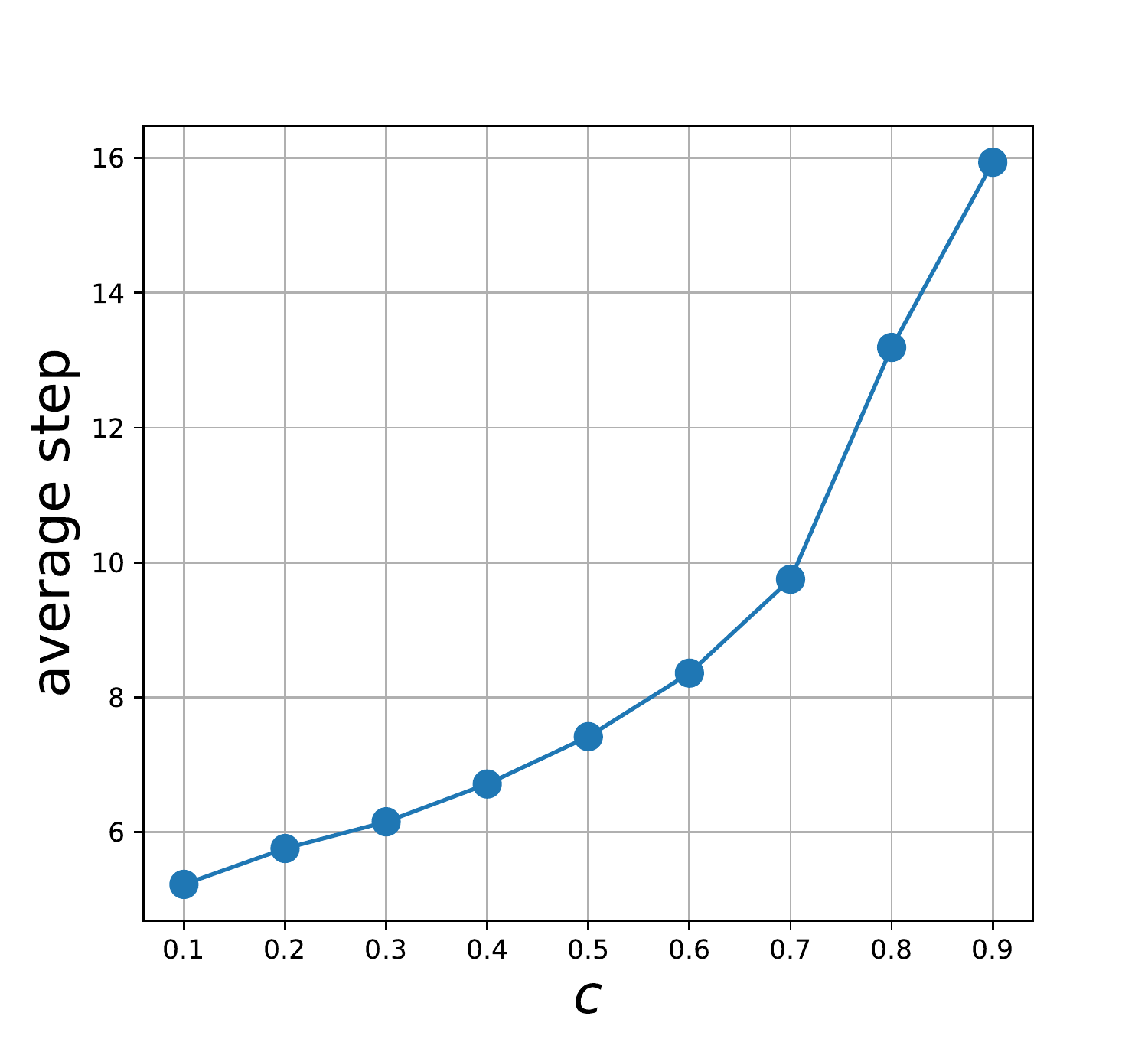}
	\end{minipage}
\caption{{\bf The $\alpha_c$ in the mutual fund and stocks network and the corresponding average step.} (a) The relationship between $\alpha_c$ and $c$. The linearity is the same with Fig.~2(a) in the main text. (b) The average step when $\alpha_c=1-c$ using the mutual fund as investors instead of mutual fund companies. It is obvious that average step are higher as the mutual fund and stocks network is less denser than the network of mutual fund companies and stocks so it would take longer for the contagion to stop.} 	
\label{fig:alpha_c_fund}
\end{figure}

\newpage
\begin{figure}[h!]
	\begin{minipage}{0.5\linewidth}
		\centering
		\includegraphics[width =0.95\linewidth]{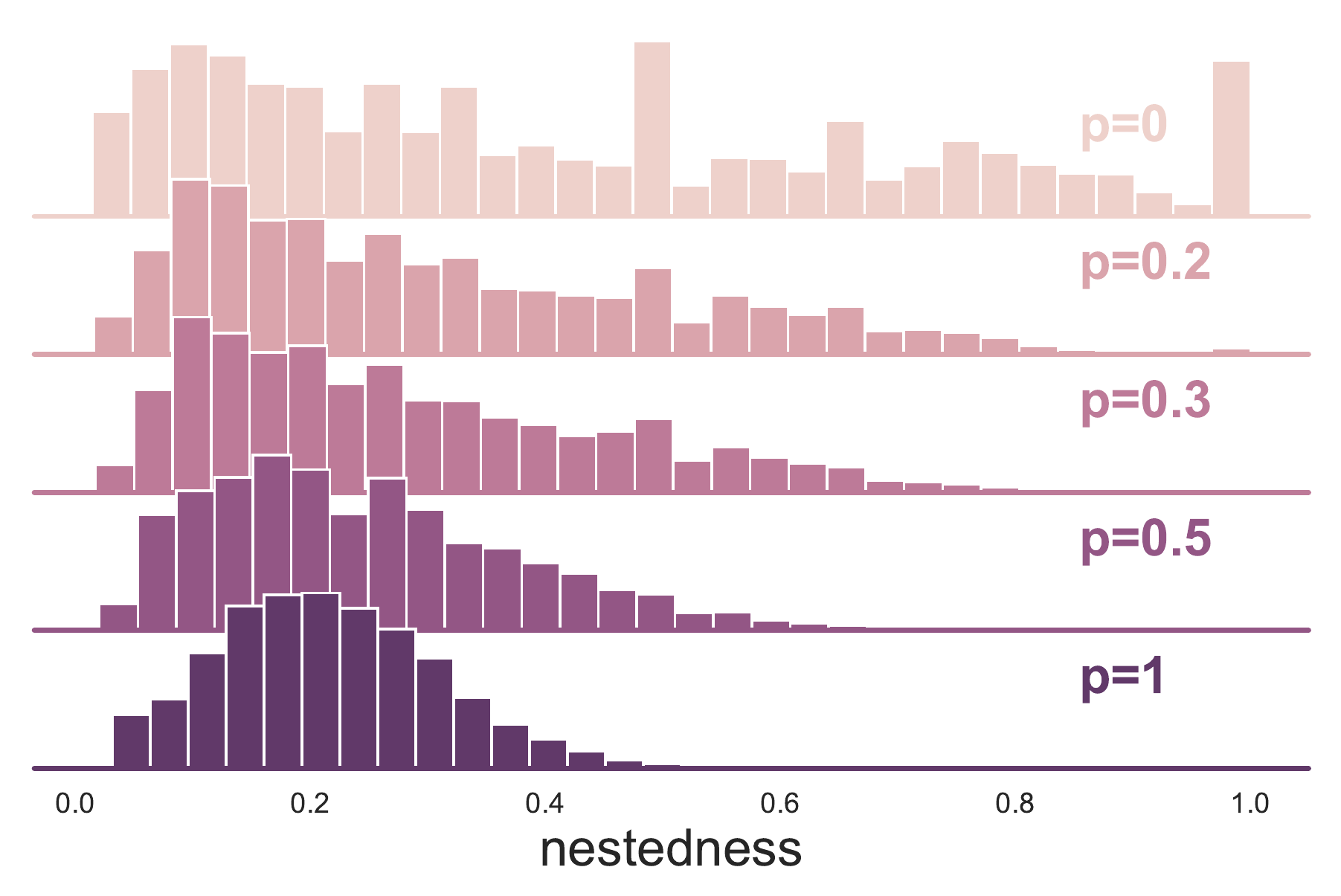}
	\end{minipage}
	\begin{minipage}{0.5\linewidth}
		\centering
		\includegraphics[width =0.95\linewidth]{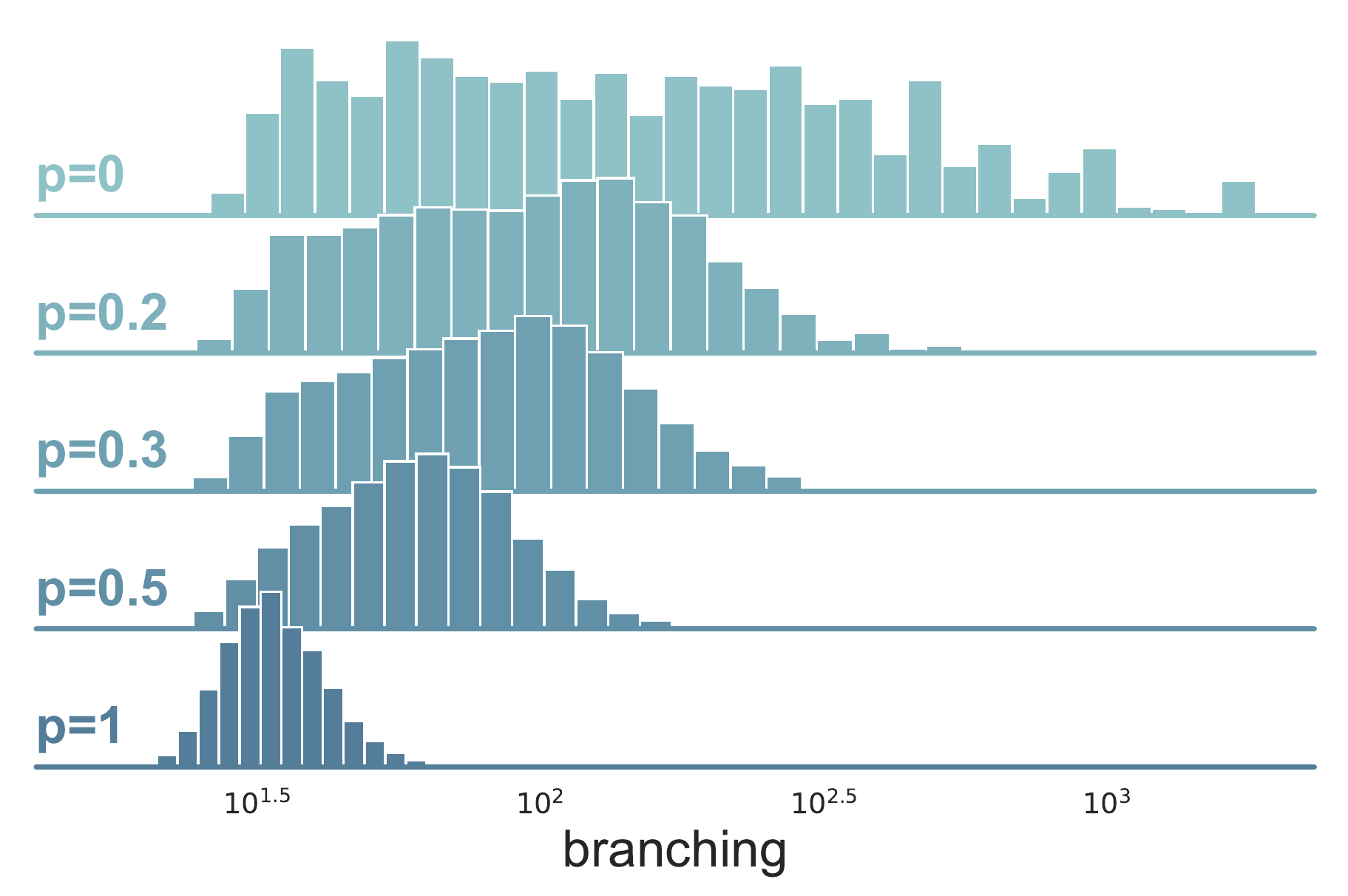}
	\end{minipage}
	
\caption{{\bf The distribution of nestedness and branching when randomizing a proportion of $p$ edges on the original network.} The $p$ denotes the proportion of which the edges in the original network are randomized, which is the same with Fig.~6 in the main text. Each of the histogram is based on one round of randomization experiment for every $p$. For visualization purpose, the histograms do not share the same y-axis. It can be seen that distributions of both the nestedness and branching have been changed when partially or fully randomizing the original network. } \label{fig:random_nest_branch}
\end{figure}

\end{appendices}

\end{document}